\documentclass[twocolumn,aps,showpacs,prb,tightenlines,amsmath,amssymb,superscriptaddress]{revtex4}
\usepackage{graphicx}
\usepackage{amssymb}
\usepackage{amsmath}
\usepackage{colordvi}

\begin{document}

\title{Kinetics of spin coherence of electrons  in $n$-type InAs 
quantum wells under intense terahertz laser fields}

\author{J. H. Jiang}
\affiliation{Hefei National Laboratory for Physical Sciences at
  Microscale, University of Science and Technology of China, Hefei,
  Anhui, 230026, China}
\affiliation{Department of Physics,
  University of Science and Technology of China, Hefei,
  Anhui, 230026, China}
\author{M. W. Wu}
\thanks{Author to whom correspondence should be addressed}
\email{mwwu@ustc.edu.cn.}
\affiliation{Hefei National Laboratory for Physical Sciences at
  Microscale, University of Science and Technology of China, Hefei,
  Anhui, 230026, China}
\affiliation{Department of Physics,
  University of Science and Technology of China, Hefei,
  Anhui, 230026, China}
\altaffiliation{Mailing Address}
\author{Y. Zhou}
\affiliation{Hefei National Laboratory for Physical Sciences at
  Microscale, University of Science and Technology of China, Hefei,
  Anhui, 230026, China}

\date{\today}

\begin{abstract}
Spin kinetics in $n$-type InAs quantum wells under intense terahertz
laser fields is investigated by developing fully microscopic kinetic
spin Bloch equations via the Floquet-Markov theory and the nonequilibrium
Green's function approach, with all the relevant
scattering, such as the electron-impurity, electron-phonon, and
electron-electron Coulomb scattering  explicitly included.
We find that a {\em finite} steady-state terahertz  spin polarization
induced by the  terahertz laser field,
first predicted by Cheng and Wu [Appl. Phys. Lett. {\bf 86}, 032107
(2005)] in the absence of dissipation, exists
even in the presence of all the scattering. We further
discuss the effects of the terahertz laser fields on the spin
relaxation and the  steady-state spin
polarization. It is found that the terahertz laser fields can {\em strongly}
affect the spin relaxation via hot-electron effect
 and the terahertz-field-induced effective magnetic
field in the presence of spin-orbit
coupling. The two effects compete with each other, giving rise to
{\em non-monotonic} dependence of the spin relaxation time as well as the
amplitude of the steady state spin polarization on
the terahertz field strength and
frequency. The terahertz field dependences of these quantities are
investigated for various impurity densities, lattice temperatures, and
strengths of the spin-orbit coupling. Finally, the importance of the
electron-electron Coulomb scattering on spin kinetics is also addressed.

\end{abstract}

\maketitle

\section{Introduction}

Generating and manipulating  spin coherence of electrons is one of the most
important research focuses of semiconductor spintronics
community.\cite{spintronics,wolf,dot}
 There have been many proposals to use
electric field rather than magnetic field to generate and
manipulate electron spin
coherence.\cite{Salis,Tokura,Rashba0,Rashba1,Rashba2,Cheng,Rashba3,Duckheim,yzhou%
,Rodriguez,Jiang,Golovach,Fabian,Bulaev,wu-trans-pulse,Pershin}
The mechanism of such proposals is that when the spin degree
of freedom is coupled to the orbital degree of freedom via spatial
varying $g$-tensor (or magnetic field) or spin-orbit coupling (SOC)
(such as the Rashba,\cite{Rashba} the
Dresselhaus\cite{Dresselhaus} and the strain-induced\cite{opt-or} SOC),
the electric field can act directly on spin through driving the orbital motion.
Recently, Kato {\it et al.} achieved coherent spin rotation via gigahertz
electric field
applied along the growth direction of the $g$-tensor engineered
GaAs/Al$_x$Ga$_{1-x}$As parabolic quantum
wells.\cite{Kato0} It has also been demonstrated experimentally that
the SOC can enable electrical
control of spin coherence without magnetic field.\cite{Kato1,Crooker,Meier}
Rashba and Efros showed that even preferable to ac magnetic fields,
ac electric fields can efficiently induce
spin resonance in the presence of SOC in quantum
wells, especially when the ac electric field is the
in-plane one.\cite{Rashba0,Rashba1,Rashba2,Rashba3} This effect is called
electric dipole spin resonance which was later observed by Kato
{\it et al.} in bulk GaAs,\cite{Kato1}  Meier {\it et al.} in
GaAs/InGaAs quantum wells,\cite{Meier} and Nowack
  {\it et al.} in GaAs quantum dots.\cite{Nowack}
In these investigations, only weak
electric fields are applied.

Recently, Cheng and Wu showed theoretically that in InAs quantum wells, a
strong in-plane terahertz (THz) electric field ($\sim1$ kV/cm) can induce a large
spin polarization ($\sim 10$\ \%) oscillating at the same
frequency of the THz driving field when dissipations are not
considered.\cite{Cheng} This indicates that using strong THz electric field is a
promising way to achieve high-frequency spin manipulation and spin
generation in InAs-based nanostructures where the spin splitting
  is of the order of THz.\cite{Grundler,Sato} They also showed
that the strong THz field can greatly modify the density of states
 via the dynamical Franz-Keldysh effect,\cite{Jauho} the sideband
effect\cite{sideband} and the ac Stark
effect.\cite{ACStark,ACReview1,ACReview2,Ganichevbook} Later,
Jiang {\em et al.} predicted similar effects in singly charged InAs
quantum dots.\cite{Jiang}
As the density of states of the electron spin system
is greatly modified by the intense THz fields, the dissipation
  effects may also be manipulated. However, up till now there is  few
study on the dissipative kinetics of  strongly-driven electron
  spin system, especially from a fully microscopic approach.
Previously, we have demonstrated that
intense THz driving field in InAs quantum dots can elongate spin
relaxation time (SRT) by more than one order of magnitude.\cite{Jiang2}
The underlying physics is that the sideband effect strongly
modulates the phonon-induced spin-flip transition rates. The effects
of intense THz fields on spin relaxation in two dimensional electron
system (2DES) are still unknown. The spin relaxation mechanism in 2DES
is quite different from that in quantum dots. In the
driving-field-free limit,
it is widely accepted that spin relaxation in 2DES is dominated by
D'yakonov-Perel' (DP) mechanism.\cite{spintronics,opt-or}
 Previously, spin
relaxation and spin dephasing
have been closely studied using the kinetic spin Bloch equation approach
 developed by Wu {\em et al.}\cite{wu-review} in intrinsic,
$n$-type and $p$-type semiconductors, in both
Markovian and non-Markovian limits, even in systems far away from
the equilibrium (under strong static electric field or with high spin
polarization).\cite{wu-early,wu-later,wu-hole,wu-highP,wu-hot-e,wu-helix,wu-lowT,wu-nonM,wu-bap}
The theory, which includes all relevant scattering (such as
the electron-impurity,  electron-phonon and  electron-electron Coulomb
scattering) explicitly,
agrees very well with experiments.\cite{wu-lowT,wu-Exp-Aniso}
Many predictions from the theory have recently been confirmed
experimentally.\cite{david,wu-Exp-HighP}

In this work, we first extend the theory to study the spin kinetics under
intense THz laser fields in InAs quantum wells. The kinetic spin
 Bloch equations are
derived in the spirit of the Floquet-Markov approach\cite{FM} via
nonequilibrium Green function method.\cite{Haugbook} The
Floquet-Markov approach combines the Floquet theory, which solves
the time-dependent Schr\"odinger equation of the strongly-driven
system {\em non-perturbatively},\cite{Shirley,ACReview1} with the
Born-Markov approximation which is widely used in the derivation of
the equation of motion for the reduced density matrix of the
concerned system.\cite{QN} The theory is frequently applied in the
study of dissipative dynamics of strongly-driven
systems.\cite{ACReview1,ACReview2,Jiang2} With the extended kinetic
spin Bloch equations, we are able to investigate the effect of the
strong THz fields on spin kinetics. We show that the steady-state
THz spin polarization induced by the THz laser field, first
predicted by Cheng and Wu in the dissipation-free case,\cite{Cheng} still
{\em exists} in the presence of the full dissipation. Moreover, we
investigate how this spin polarization as well
as the spin relaxation are manipulated by the external
THz laser fields under various conditions. The predicted spin dynamics can
be readily confirmed by Faraday and/or Kerr rotation
measurements\cite{FR,KR} under intense THz irradiation.

The paper is organized as follows: In Sec.\ II, we set up the model and
establish the kinetic spin Bloch equations. In Sec.\ III, we present
our numerical results. We conclude and discuss in Sec.\ IV.

\section{Model and Formalism}

\subsection{Model}

We consider a 2DES confined in InAs quantum well. The
confinement along the $z$ direction is so strong (well width $a=5$~nm)
that only the lowest subband is considered. The THz electric
field ${\bf E}_{THz}(t) = {\bf E}\sin(\Omega t)$ is applied in
the quantum well plane. Here $\Omega=2\pi\nu$ is the angular
  velocity, with $\nu$ being the frequency of the THz field.
In experiments, the THz
field can be provided by the free electron laser.\cite{Ramian}
In the Coulomb gauge, the vector
potential is ${\bf A}(t) = {\bf E}\cos(\Omega t)/\Omega$ and the
scalar potential is $\Phi = 0$.
In InAs quantum wells, the dominant SOC is the Rashba SOC.\cite{Rashba}
As the Rashba SOC is
rotational invariant, we take  ${\bf  E}_{THz}$ along $x$ axis. The
total Hamiltonian is then
\begin{equation}
  H=H_{e} + H_{ei} + H_{ee} + H_{ep} + H_{ph}.
\end{equation}
Here $H_{ph}=\sum_{\lambda,{\bf Q}} \omega_{\lambda,{\bf Q}}
\hat{a}_{\lambda,{\bf Q}}^{\dagger} \hat{a}_{\lambda,{\bf Q}}$
represents the phonon Hamiltonian (We take $\hbar\equiv 1$ throughout the paper).
 $H_{ei}=\sum_{{\bf k}{\bf Q},\sigma,j}U_{{\bf Q}} I(iq_z)
  e^{-i{\bf Q}\cdot {\bf R}_{j}} \hat{\sl c}^{\dagger}_{{\bf k}\sigma} \hat{\sl
  c}_{{\bf k}-{\bf q}\sigma}$,
$H_{ee}=\frac{1}{2}\sum_{{\bf k}{\bf
  k}^{\prime}{\bf Q},\sigma\sigma^{\prime}} V_{{\bf Q}} |I(iq_z)|^2 \hat{\sl c}^{\dagger}_{{\bf
  k}\sigma} \hat{\sl c}^{\dagger}_{{\bf k}^{\prime}\sigma^{\prime}}\hat{\sl
c}_{{\bf k}^{\prime}-{\bf q}\sigma^{\prime}} \hat{\sl c}_{{\bf
  k}+{\bf q}\sigma}$ and $H_{ep}= \sum_{\lambda,{\bf Q}{\bf k}}
  M_{\lambda,{\bf Q}} I(iq_z)
(\hat{a}_{\lambda,{\bf Q}}+\hat{a}_{\lambda,{\bf -Q}}^{\dagger})\hat{\sl c}^{\dagger}_{{\bf k}\sigma}
\hat{\sl  c}_{{\bf k}-{\bf q}\sigma}$ denote the electron-impurity,
electron-electron and electron-phonon  interactions, respectively.
Here, $\omega_{\lambda,{\bf Q}}$ is the phonon
  frequency, $\hat{a}_{\lambda,{\bf Q}}$ ($\hat{\sl c}_{{\bf
  k}\sigma}$) is the phonon (electron) annihilation operator with
$\lambda$ being the phonon branch index  ($\sigma$ denoting the
electron spin index), ${\bf R}_{j}$ stands for the position of $j$th
  impurity, ${\bf Q}=({\bf q}, q_z)$ is the three-dimensional
  momentum whereas ${\bf q}$ and ${\bf k}$'s are the two-dimensional ones along
  the well plane. $I(iq_z)=(\frac{2\pi}{a})^2(e^{iq_z
  a}-1)/\{iq_z a[(\frac{2\pi}{a})^2-q_z^2]\}$ is the form factor of the lowest
subband. The matrix elements $V_{{\bf Q}}$,
$U_{{\bf Q}}$ for electron-electron and electron-impurity interactions
 as well as  $M_{\lambda,{\bf Q}}$ for the
  electron--longitudinal-optical-phonon (electron--LO-phonon)
interaction can be found in
Ref.\ \onlinecite{wu-hot-e} and the ones for the
electron--acoustic-phonon interaction can be found in
  Ref.\ \onlinecite{wu-lowT}. We
apply the random phase approximation in the screening of the Coulomb
potential.\cite{wu-lowT}

The electron Hamiltonian can be written as
\begin{equation}
  H_{e} = \sum_{{\bf k}\sigma\sigma^{\prime}}
  H_0^{\sigma\sigma^{\prime}}({\bf k},t) \hat{\sl c}^{\dagger}_{{\bf
      k}\sigma} \hat{\sl c}_{{\bf k}\sigma^{\prime}},
\end{equation}
where
\begin{eqnarray}
 && \hspace{-0.6cm}\hat{H}_0({\bf k},t)=\frac{({\bf k} +e{\bf A}(t))^2}{2m^{\ast}} \mbox{\boldmath
    $\hat{1}$\unboldmath} +
  \alpha_R\Big[\hat{\sigma}_xk_y\nonumber\\
&&\hspace{1.5cm}\mbox{} -\hat{\sigma}_y(k_x+eA(t))\Big]
  \nonumber\\ && =
\Big[\varepsilon_{\bf k} +\gamma_E
  k_x\Omega\cos(\Omega t) + E_{em}(1+\cos(2\Omega t)) \Big] \mbox{\boldmath
    $\hat{1}$\unboldmath} \nonumber\\ && \mbox{} +
  \alpha_R(\hat{\sigma}_xk_y-\hat{\sigma}_yk_x) -
  \alpha_R\hat{\sigma}_yeE\cos(\Omega t)/\Omega.
\label{hamilton}
\end{eqnarray}
Here $\varepsilon_{\bf k}=\frac{{\bf k}^2}{2m^{\ast}}$,
$\gamma_{E}=\frac{eE}{m^{\ast}\Omega^2}$ and
$E_{em}=\frac{e^2E^2}{4m^{\ast}\Omega^2}$. It is noted that the last
term manifests that the THz electric field acts as a THz magnetic field
along the $y$ axis
\begin{equation}
  B_{\mbox{eff}}= 2\alpha_R eE\cos(\Omega t)/\left(|g|\mu_B\Omega\right),
\label{beff}
\end{equation}
where $g$ is the electron $g$-factor. We will show later that this
THz-field-induced
effective magnetic field has many important effects on  spin
kinetics. The term proportional to $E_{em}$ is responsible for the
dynamical Franz-Keldysh effect.\cite{Jauho,Cheng} This term does
  not contain any dynamic variable of the electron system and thus
  has no effect on the kinetics of the electron system.
Usually, the largest time-periodic term is the term $\gamma_E
k_x\Omega\cos(\Omega t)$, where the sideband effect mainly comes from.
Under an intense THz field, this term can be comparable to or larger than
 $\varepsilon_{\bf k}$. It should be noted that this anisotropic
term breaks down $k_x\to-k_x$ symmetry of the Hamiltonian. We
will show later that this asymmetry leads to nonzero value of the
average of $k_x$ over the electron system when the momentum scattering
is included.

The Schr\"odinger equation for electron with momentum ${\bf k}$ is
\begin{equation}
i\partial_{t}\Psi_{{\bf k}}(t) = \hat{H}_0({\bf k},t) \Psi_{{\bf
    k}}(t).
\end{equation}
According to the Floquet theory,\cite{Shirley} the solution to the above
equation reads
\begin{eqnarray}
 \Psi_{{\bf k}\eta}(t)
&=& e^{i{\bf k}\cdot{\bf r}-i\varepsilon_{\bf k}t}\phi_{1}(z)
\xi_{{\bf k}\eta}(t) \nonumber \\ &&\mbox{} \times
e^{-i[\gamma_E k_x \sin(\Omega t) + E_{em}t+E_{em}
\frac{\sin(2\Omega t)}{2\Omega}]},
\label{Floquet}
\end{eqnarray}
with $\eta=\pm$ denoting the spin branch and $\phi_{1}(z)$ being the
wavefunction of the lowest
subband. $\xi_{{\bf k}\eta}(t)=e^{-iy_{{\bf
    k}\eta}t}\sum_{n\sigma}\upsilon_{n\sigma}^{{\bf
    k}\eta}e^{in\Omega t}\chi_{\sigma}$
where $y_{{\bf k}\eta}$ and $\upsilon_{n\sigma}^{{\bf k}\eta}$
are the eigen-value and eigen-vector of the  equation
\begin{eqnarray}
 (y_{{\bf k}\eta}-n\Omega) \upsilon_{n\sigma}^{{\bf k}\eta} &=&
   \frac{i\sigma}{2\Omega}\alpha_R eE (\upsilon_{n-1,-\sigma}^{{\bf
      k}\eta} +\upsilon_{n+1,-\sigma}^{{\bf k}\eta}) \nonumber\\
&&\mbox{} + \alpha_R(k_y+i\sigma k_x)\upsilon_{n,-\sigma}^{{\bf k}\eta}.
\label{floquet}
\end{eqnarray}
This equation is equivalent to Eq.\ (2) in Ref.\ \onlinecite{Cheng}.
For each ${\bf k}$, the spinors \{$|\xi_{{\bf k}\eta}(t)\rangle$\} at
any time $t$ form a complete-orthogonal basis of the spin
space.\cite{ACReview1,ACReview2,Jiang2} The time evolution operator
for state ${\bf k}$ can be written as
\begin{eqnarray}
\hat{U}_0^e({\bf
  k},t,0)&=&\sum_{\eta}|\xi_{{\bf k}\eta}(t)\rangle\langle \xi_{{\bf
    k}\eta}(0)|e^{-i[\varepsilon_{\bf k}t+\gamma_E k_x \sin(\Omega t)]}\nonumber\\
&&\mbox{}\times e^{-i[E_{em}t+E_{em}\sin(2\Omega t)/(2\Omega)]}.
\end{eqnarray}

\subsection{Kinetic spin Bloch equations}
The kinetic spin Bloch equations offer a fully microscopic way to study
spin dynamics in semiconductors, even in system with large static electric
field where the hot-electron effect is important.\cite{wu-hot-e,wu-lowT}
The electric field dependence of spin dephasing time in such system was
studied first by Weng {\em et al.}\cite{wu-hot-e} for
high temperature case and then by Zhou {\em et al.}\cite{wu-lowT}
for low temperature case. In these works, the
electric field appears only in the driving term. However, in the
case of a strong time-periodic field, studies have shown that
including this field only in the driving term is
insufficient.\cite{FM} The correct way is to evaluate the collision
integral with wavefunctions which are the solutions of the
time-dependent Schr\"odinger equation, i.e., the Floquet
wavefunctions, instead of the eigen wavefunctions in the field-free
limit.\cite{FM,Haugbook} Moreover, the Markovian approximation should be made
with respect to the spectrum determined by the Floquet
wavefunctions.\cite{FM} These improvements
constitute the Floquet-Markov theory.\cite{FM,ACReview1} Generally,
this theory works well when the driven system is in dynamically
stable regime and the system-reservoir coupling can be treated
perturbatively. Besides giving good results, this approach has the
advantage of being easy to handle, compared with the rather complicated
path-integral approach,\cite{FM} which makes it a useful approach in the study of
spin kinetics under strong time-periodic fields.
In this work, we incorporate the Floquet-Markov approach in setting up the
kinetic spin Bloch equations.
By making the Markov approximation with respect to the spectrum
determined by the Floquet states, we first establish the kinetic
equations for the single particle density operator. We then use the
Floquet states as basis functions to expand the kinetic equations and
obtain the kinetic spin Bloch equations in the presence of the strong THz
field. A similar approach has been applied to study the spin
relaxation in singly charged quantum dots under
intense THz driving fields in our recent work.\cite{Jiang2}

The kinetic spin Bloch equations for the single particle density operator
can be written as\cite{wu-review}
\begin{equation}
\label{KSBE}
  \partial_{t}\hat{\rho}_{{\bf k}}(t) =
  \partial_{t}\hat{\rho}_{{\bf k}}(t)|_{coh} +
  \partial_{t}\hat{\rho}_{{\bf k}}(t)|_{scat},
\end{equation}
where $\partial_{t}\hat{\rho}_{{\bf k}}(t)|_{coh}$ and
$\partial_{t}\hat{\rho}_{{\bf k}}(t)|_{scat}$ are  the coherent and scattering terms
respectively.
$\hat{\rho}_{\bf k}(t) =\sum_{\eta_1\eta_2}
\mbox{Tr}\Big\{\hat{c}^{\dagger}_{{\bf k}\eta_2}\hat{c}_{{\bf
    k}\eta_1}\hat{\rho}^e(t)\Big\}|\eta_1\rangle\langle \eta_2|$
represent the $2\times2$ single-particle density operators, with
$\hat{\rho}^e(t)$ and $\{|\eta\rangle\}$ denoting
the density operator of the electron system and a
complete-orthogonal basis in spin space separately.
The explicit form of the equations without the intense driving field
can be found in the work of Cheng and Wu.\cite{wu-helix}
The coherent terms, which describe the coherent precession determined
by the electron Hamiltonian $H_e$ and the Hartree-Fock
contribution of the electron--electron Coulomb interaction, can be written as
\begin{equation}
  \partial_{t}\hat{\rho}_{{\bf k}}(t)|_{coh} = -i[\hat{H}_0({\bf
    k},t), \hat{\rho}_{{\bf k}}(t)] - i[\hat{\Sigma}^{HF}({\bf k},t),
  \hat{\rho}_{{\bf k}}(t)].
\end{equation}
Here $\hat{\Sigma}^{HF}({\bf k},t)=-\sum_{{\bf
    k}^{\prime},q_z}V_{{{\bf k}-{\bf
      k}^{\prime}},q_z}|I(iq_z)|^2\hat{\rho}_{{\bf k}^{\prime}}(t)$ is the Coulomb
 Hartree-Fock self-energy.
The scattering terms are
composed of terms due to the electron-impurity ($\partial_{t}\rho_{{\bf k}}|_{ei}$), electron-phonon
($\partial_{t}\rho_{{\bf k}}|_{ep}$), and
electron-electron ($\partial_{t}\rho_{{\bf k}}|_{ee}$) scattering respectively.
In the interaction picture, or the ``Floquet picture'',
\begin{equation}
  \hat{\rho}^{F}_{\bf k}(t) = \hat{U}_0^{e\ \dagger}({\bf k},t,0)
  \hat{\rho}_{\bf k}(t) \hat{U}_0^{e}({\bf k},t,0),
\end{equation}
and under the generalized Kadanoff-Baym Ansatz,\cite{Haugbook}
these scattering terms read
\begin{widetext}
\begin{eqnarray}
 \left. \partial_{t}\rho^{F(\eta\eta^{\prime})}_{{\bf k}}(t)\right|_{ei}
 &=& -
\sum_{{\bf k}^{\prime},q_z,n,\eta_1\eta_2\eta_3} \pi n_i U^2_{{\bf k}-{\bf
      k}^{\prime},q_z} |I(iq_z)|^2  \bigg{[} \Big{\{} S^{(\eta\eta_1)}_{{\bf k},{\bf
          k}^{\prime}}(t,0)S^{(n)(\eta_2\eta_3)}_{{\bf
      k}^{\prime},{\bf k}} \delta(n\Omega+\bar{\varepsilon}_{{\bf
      k}^{\prime}\eta_2}-\bar{\varepsilon}_{{\bf k}\eta_3} ) \nonumber \\
 && \hspace{1.cm} \mbox{} \times \Big{(}\rho^{>F(\eta_1\eta_2)}_{{\bf
 k}^{\prime}}(t) \rho^{<F(\eta_3\eta^{\prime})}_{{\bf k}} (t)
  - \rho^{<F(\eta_1\eta_2)}_{{\bf
 k}^{\prime}}(t) \rho^{>F(\eta_3\eta^{\prime})}_{{\bf k}}(t) \Big{)} \Big{\}}
  +\Big{\{}\eta\leftrightarrow\eta^{\prime}\Big{\}}^{\ast} \bigg{]},
\label{scatei}
\end{eqnarray}
\begin{eqnarray}
 \left. \partial_{t}\rho^{F(\eta\eta^{\prime})}_{{\bf k}}(t)\right|_{ep}
&=& -{\sum_{{\bf k}^{\prime},q_z,n,\lambda,\pm,\eta_1\eta_2\eta_3} \pi
  |M_{\lambda, {\bf k}-{\bf k}^{\prime},q_z}|^2 |I(iq_z)|^2 }
  \bigg{[} \Big{\{}
  S^{(\eta\eta_1)}_{{\bf k},{\bf k}^{\prime}}(t,0)
  S^{(n)(\eta_2\eta_3)}_{{\bf k}^{\prime},{\bf k}}
  e^{\mp it\omega_{\lambda, {\bf k}-{\bf k}^{\prime},q_z}}
   \nonumber \\
&& \hspace{1.cm} \mbox{} \times \delta(\pm\omega_{\lambda, {\bf k}-{\bf
      k}^{\prime},q_z}+n\Omega + \bar{\varepsilon}_{{\bf
      k}^{\prime}\eta_2} - \bar{\varepsilon}_{{\bf k}\eta_3} ) \Big{(}
   N^{\pm}_{\lambda, {\bf k}-{\bf k}^{\prime},q_z}
   \rho^{>F(\eta_1\eta_2)}_{{\bf k}^{\prime}}(t)
   \rho^{<F(\eta_3\eta^{\prime})}_{\bf k}(t) \nonumber \\
&& \hspace{1.5cm} \mbox{}
  - N^{\mp}_{\lambda, {\bf k}-{\bf k}^{\prime},q_z} \rho^{<F(\eta_1\eta_2)}_{{\bf
       k}^{\prime}}(t) \rho^{>F(\eta_3\eta^{\prime})}_{\bf k}(t)
   \Big{)} \Big{\}} +
{\Big{\{}\eta\leftrightarrow\eta^{\prime}\Big{\}}^{\ast}} \bigg{]},
\label{scatep}
\end{eqnarray}
\begin{eqnarray}
\left.\partial_{t}\rho^{F(\eta\eta^{\prime})}_{{\bf k}}(t)\right|_{ee}
&=& -  \sum_{{\bf k}^{\prime},{\bf k}^{\prime\prime},n,n^{\prime}}\
  \sum_{\eta_1...\eta_7} \pi \Big{[}\sum_{q_z} V_{{\bf k}-{\bf k}^{\prime},q_z}
  |I(iq_z)|^2 \Big{]}^2
  \bigg{[} \Big{\{}
    T^{(\eta\eta_1)}_{{\bf k},{\bf k}^{\prime}}(t,0)
    T^{(n^{\prime})(\eta_2\eta_3)}_{{\bf k}^{\prime},{\bf k}}
    T^{(n-n^{\prime})(\eta_4\eta_5)}_{{\bf k}^{\prime\prime},{\bf
        k}^{\prime\prime}-{\bf k}+{\bf k}^{\prime}} T^{(\eta_6\eta_7)}_{{\bf
        k}^{\prime\prime}-{\bf k}+{\bf k}^{\prime},{\bf
    k}^{\prime\prime}}(t,0)
\nonumber \\
&& \hspace{0.cm} \mbox{} \times \delta( n\Omega+
\bar{\varepsilon}_{{\bf k}^{\prime}\eta_2}-\bar{\varepsilon}_{{\bf
    k}\eta_3}+\bar{\varepsilon}_{{\bf k}^{\prime\prime}\eta_4}
    -\bar{\varepsilon}_{{\bf k}^{\prime\prime}-{\bf k}+{\bf
    k}^{\prime}\eta_5} )
     \Big{(} \rho^{>F(\eta_1\eta_2)}_{{\bf k}^{\prime}}(t)
\rho^{<F(\eta_3\eta^{\prime})}_{\bf k}(t)
\rho^{<F(\eta_5\eta_6)}_{{\bf
        k}^{\prime\prime}-{\bf k}+{\bf k}^{\prime}}(t)
\rho^{>F(\eta_7\eta_4)}_{{\bf k}^{\prime\prime}}(t) \nonumber \\
&&\hspace{0.cm}\mbox{}
-   \rho^{<F(\eta_1\eta_2)}_{{\bf k}^{\prime}}(t)
\rho^{>F(\eta_3\eta^{\prime})}_{\bf k}(t)
\rho^{>F(\eta_5\eta_6)}_{{\bf
        k}^{\prime\prime}-{\bf k}+{\bf k}^{\prime}}(t)
\rho^{<F(\eta_7\eta_4)}_{{\bf k}^{\prime\prime}}(t) \Big{)} \Big{\}}
    +{\Big{\{}\eta\leftrightarrow\eta^{\prime}\Big{\}}^{\ast}} \bigg{]}.
\label{scatee}
\end{eqnarray}
\end{widetext}
In these equations, $N^{\pm}_{\lambda, {\bf k}-{\bf k}^{\prime},q_z}=N_{\lambda, {\bf k}-{\bf
    k}^{\prime},q_z}+\frac{1}{2}(1 \pm 1)$ stands for the phonon number, $n_i$ is the
impurity density, $\hat{\rho}^{>}_{\bf k} = \hat{{\bf
    1}} - \hat{\rho}_{\bf k}$,
$\hat{\rho}^{<}_{\bf k} = \hat{\rho}_{\bf
    k}$,  and $\bar{\varepsilon}_{{\bf k}\eta}=\varepsilon_{{\bf
    k}} + y_{{\bf k}\eta}$.
\begin{eqnarray}
 S^{(\eta_1\eta_2)}_{{\bf k}^{\prime},{\bf
      k}}(t,0)&=&\langle \xi_{{\bf
    k}^{\prime}\eta_1}(t)|\xi_{{\bf k}\eta_2}(t)\rangle \nonumber \\
&& \mbox{} \times
e^{i[(\varepsilon_{{\bf k}^{\prime}}-\varepsilon_{\bf
      k})t+\gamma_E\sin(\Omega t)(k_x^{\prime}-k_x)]} \nonumber \\
 &=& \sum_{n} S^{(n)(\eta_1\eta_2)}_{{\bf
    k}^{\prime},{\bf k}}
e^{i t(n\Omega+\bar{\varepsilon}_{{\bf
      k}^{\prime}\eta_1}-\bar{\varepsilon}_{{\bf k}\eta_2}) },
\end{eqnarray}
with
\begin{equation}
S^{(n)(\eta_1\eta_2)}_{{\bf k}^{\prime},{\bf k}} =
\sum_{m\sigma}F^{{\bf k}^{\prime}\eta_1 \ast}_{m\ \sigma}\ F^{{\bf k}\eta_2}_{n+m\ \sigma}.
\end{equation}
Here $F^{{\bf k}\eta}_{n\ \sigma}=\sum_{m} \upsilon^{{\bf
    k}\eta}_{n+m\ \sigma} J_{m}(\gamma_E k_x)$ with $J_{m}(x)$ standing
    for the $m$th order Bessel function.
\begin{eqnarray}
 T^{(\eta_1\eta_2)}_{{\bf k}^{\prime},{\bf k}}(t,0)
 &=&\langle \xi_{{\bf
    k}^{\prime}\eta_1}(t)|\xi_{{\bf k}\eta_2}(t)\rangle
e^{i(\varepsilon_{{\bf k}^{\prime}}-\varepsilon_{\bf k})t}  \nonumber \\
 &=& \sum_{n} T^{(n)(\eta_1\eta_2)}_{{\bf
    k}^{\prime},{\bf k}}
e^{i t(n\Omega+\bar{\varepsilon}_{{\bf
      k}^{\prime}\eta_1}-\bar{\varepsilon}_{{\bf k}\eta_2}) },
\end{eqnarray}
with
\begin{equation}
T^{(n)(\eta_1\eta_2)}_{{\bf k}^{\prime},{\bf k}} =
\sum_{m\sigma}\upsilon^{{\bf k}^{\prime}\eta_1 \ast}_{m\ \sigma}\
\upsilon^{{\bf k}\eta_2}_{n+m\ \sigma}.
\end{equation}
$\{\eta\leftrightarrow\eta^{\prime}\}$ stands for the
same terms as in the previous $\{\}$ but with the interchange
$\eta\leftrightarrow\eta^{\prime}$. The term of the
electron-electron scattering is
quite different from those of the electron-impurity  and
electron-phonon scattering, as the momentum
  conservation eliminates the term of  $e^{i\gamma_E\sin(\Omega t)k_x}$.
The coherent term in the Floquet picture reads
\begin{eqnarray}
\left. \partial_{t}\hat{\rho}^{F}_{{\bf k}}(t)\right|_{coh} &=&
i\sum_{{\bf
      k}^{\prime},q_z} V_{{\bf k}-{\bf
      k}^{\prime},q_z}|I(iq_z)|^2  \nonumber \\
 && \hspace{-0.8cm} \mbox{} \times \big{[}\hat{S}_{{\bf k},{\bf
      k}^{\prime}}(t,0)\hat{\rho}^{F}_{{\bf k}^{\prime}}(t)
    \hat{S}_{{\bf k}^{\prime},{\bf k}}(t,0),\ \hat{\rho}^{F}_{{\bf
    k}}(t)\big{]}.
\end{eqnarray}
At zero THz field, the
sideband summations are omitted and the above equations go back to
those in Ref.\ \onlinecite{wu-helix}.
In Appendix\ A, we use the electron-impurity scattering as an example to show how to
derive the scattering terms in the Floquet-Markov limit.

The above equations clearly show the sideband effects, i.e., $n\Omega$
in the $\delta$-functions. The extra energy, $n\Omega$, is provided by
the THz field during each scattering process. This makes transitions
from the low-energy states (small $k$) to high-energy ones (large
$k$) become possible, even through the elastic electron-impurity
scattering. These processes are the sideband-modulated scattering
processes. For example, the weight of the $n$-th sideband-modulated
 electron-impurity scattering, $|S^{(n)(\eta_2\eta_3)}_{{\bf
    k}^{\prime},{\bf k}}|$, is approximately
  $\delta_{\eta_2\eta_3}|J_{n}(\gamma_E(k_x-k_x^{\prime}))|$
when $\gamma_E k_x\Omega\cos(\Omega t)$ in the Hamiltonian
[Eq.\ (\ref{hamilton})] is the main source of the sideband effect. This
term is important when $n$ is around $\pm N_{m}$, with $N_{m}$ representing
the integer part of $\gamma_E(k_x-k_x^{\prime})$. In fact, the sideband-modulated
scattering makes the electron distribution in the three energy
  ranges around $\varepsilon_{{\bf k}}$, $\varepsilon_{{\bf k}}\pm
  N_{m}\Omega$ tend to be more uniform according to Eq.\ (\ref{scatei}).
This, together with the other two scatterings,
lead to the thermalization of the electron system,
i.e., the hot-electron effect. Consequently, the electron temperature
$T_e$ becomes larger than the lattice temperature $T$.
Previously, it has been found that the hot-electron effect has large
influence on spin dephasing and spin relaxation under high static
electric field.\cite{wu-hot-e,wu-lowT} In this paper, we will show
similar effects in the case with THz driving field.
Finally, it is noted that, as the sideband effect mainly comes from
the term of $\gamma_E k_x\Omega\cos(\Omega t)$, the electron-impurity
and electron-phonon scattering plays the leading role in transferring
energy from the THz electric field to the electron system.

A pronounced feature of the kinetic equations is that all the
scattering terms are directly time-dependent.
In our previous study on spin dynamics in quantum dots with strong
THz field,\cite{Jiang2} due to the fact that the spin-flip
electron-phonon scattering rates are much smaller than the Zeeman
splitting and the THz frequency, one can use the
rotating-wave-approximation (RWA) treatment of the scattering terms and
consequently only the time-independent terms are kept.\cite{FM} Here, as the
scattering rate (especially that due to the electron-electron Coulomb scattering)
 is of the same order of the THz frequency, the
RWA is no longer applicable. Thus the scattering terms
become explicitly time-dependent. Moreover, the scattering
and coherent terms  are time-periodic
functions with period $T_0=2\pi/\Omega$. Consequently the kinetic
spin Bloch equations
are time-periodic differential equations, whose eigen-modes have the
general form of $\hat{\tilde{\rho}}^{\alpha}_{\bf k}=e^{i\mu_{\bf
    k}^{\alpha}t}\sum_{n}\hat{Q}_{{\bf k}}^{\alpha,n}e^{in\Omega t}$
($\alpha=1,2,3,4$) according to Floquet theorem.\cite{ACReview2} Therefore, the
solutions of the equations can be expressed as
$\hat{\rho}_{\bf
  k}=\sum_{\alpha}C_{\alpha}\hat{\tilde{\rho}}^{\alpha}_{\bf k}$ with
$C_{\alpha}$ denoting the time-independent coefficients.

The kinetic spin Bloch equations are solved numerically with
the numerical scheme laid out in Appendix\ B. After that,
$\rho^{F(\eta\eta^{\prime})}_{{\bf k}}(t)$ for each ${\bf k}$ is
obtained. From
\begin{eqnarray}
&& \hspace{-0.5cm} \rho^{F(\eta\eta^{\prime})}_{{\bf
    k}}(t) = \langle \xi_{{\bf k}\eta}(0) |\hat{U}_0^{e\ \dagger}({\bf k},t,0)
  \hat{\rho}_{\bf k}(t) \hat{U}_0^{e}({\bf k},t,0)
  |\xi_{{\bf k}\eta^{\prime}}(0)\rangle
\nonumber \\
&& \hspace{1.05cm}  = \langle \xi_{{\bf k}\eta}(t) | \hat{\rho}_{\bf
  k}(t) |\xi_{{\bf k}\eta^{\prime}}(t)\rangle,
\end{eqnarray}
by performing an unitary transformation, one comes to
the single particle
density matrix $\hat{\rho}_{\bf k}(t)$ in the collinear basis
\{$|\sigma\rangle$\} which is composed by the
eigen-states of $\hat{\sigma}_{z}$. In this
spin space, the spin polarization along any direction can be obtained
readily, e.g., $S_z=\sum_{{\bf k}}\frac{1}{2}(\rho_{\bf
  k}^{\uparrow\uparrow}-\rho_{\bf k}^{\downarrow\downarrow})$,
$S_x=\sum_{{\bf k}}\mbox{Re}\{\rho_{\bf k}^{\uparrow\downarrow}\}$,
$S_y=-\sum_{{\bf k}}\mbox{Im}\{\rho_{\bf k}^{\uparrow\downarrow}\}$.
From the temporal evolution of $S_z$, the SRT is
extracted.

Finally we briefly comment on the gauge invariance.
Although the above formalism is derived in the Coulomb gauge, the
obtained physical observables, e.g., $S_z$, is gauge invariant.
This is because
$S_z=\mbox{Tr}(\frac{1}{2}\hat{\sigma}_{z}\hat{\rho}_{{\bf k}}) =
\sum_{{\bf k}\sigma\sigma^{\prime}}\frac{1}{2}\langle {\bf
  k}\sigma|\hat{\sigma}_{z}|{\bf k}\sigma^{\prime}\rangle\langle {\bf
  k}\sigma^{\prime}| \hat{\rho}_{{\bf k}} | {\bf
  k}\sigma\rangle$. Any gauge transformation, ${\bf A}\to {\bf A} +
\mbox{\boldmath$\nabla$\unboldmath} \chi({\bf r},t)$ and
$\Phi\to\Phi-\partial_{t}\chi({\bf r},t)$, gives $|{\bf
  k}\sigma\rangle\to e^{-ie\chi({\bf r},t)}|{\bf
  k}\sigma\rangle$. %The phase factor $e^{-ie\chi({\bf r},t)}$
%of the wavefunction cancels with each other.
Thus the results are gauge invariant.

\section{Numerical results and discussions}
We numerically solve the kinetic spin Bloch equations, Eq.\
(\ref{KSBE}), with all the scattering mechanisms explicitly included,
to study the spin kinetics in InAs quantum wells under intense THz
laser fields. The parameters used are listed in Table\ I.\cite{para}
The density of the 2DES is $N_e=10^{11}$~$\mbox{cm}^{-2}$ and the
quantum well width is $a$=5~nm throughout the paper. The Rashba
parameter and the frequency of the THz field is taken to be
$\alpha_R$=30~meV$\cdot$nm and $\nu=0.65$~THz respectively, unless
otherwise specified. The initial distribution of the electron system is
chosen to be a thermalized distribution under the THz
field, which is obtained by sufficient long-time (typically $\sim$10~ps)
evolution from a spin-polarized Fermi distribution at the lattice
temperature $T$: $\rho_{{\bf
    k}}^{\uparrow\downarrow}=0$, $\rho_{{\bf
    k}}^{\sigma\sigma}=1/\big{[}e^{(\varepsilon_{k}-\mu_{\sigma})/k_BT}+1\big{]}$
($\mu_{\sigma}$ denotes the chemical potential of electrons with spin $\sigma$)
with the SOC being turned off.\cite{wu-hot-e}

The following  of this section is divided into two parts.
In the first part, we study the spin pumping due to the THz
laser field. We first show that the THz field can pump spin
polarization, first predicted
by Cheng and Wu in the dissipation-free
case,\cite{Cheng} even in the presence of full dissipation. We then
investigate the amplitude of the steady-state spin polarization as
function of the THz field strength and frequency for various impurity
densities, lattice temperatures and Rashba SOC parameters. In the second
part, we investigate the spin dynamics with finite initial
spin polarization. We first show the temporal evolution of spin
polarization for a typical case with different THz fields. We then
study the dependence of the SRT on the strength and
frequency of THz field  under various conditions.

\begin{table}[htbp]
\caption{Parameters used in the calculation}
\begin{tabular}{lllllll}\hline\hline
D&\mbox{}&$5.9\times10^3$\ kg/m$^3$&\mbox{}\mbox{}&$v_{st}$&\mbox{}
&$1.83\times10^3$\ m/s\\
$v_{sl}$&\mbox{}&$4.28\times 10^3$\ m/s&\mbox{}\mbox{}&$e_{14}$&\mbox{}&
$0.35\times 10^9$\ V/m\\
$\varXi$&\mbox{}&5.8\ eV&\mbox{}\mbox{}&$\omega_{LO}$&\mbox{}&27.0\ meV\\
$\kappa_0$&\mbox{}&15.15&\mbox{}\mbox{}&$\kappa_{\infty}$&\mbox{}&12.25\\
$g$&\mbox{}&$-14.7$&\mbox{}\mbox{}&$m^{\ast}$&\mbox{}&$0.0239m_0$\mbox{}\\
\hline\hline
\end{tabular}
\label{tab:parameter}
\end{table}

\subsection{Spin pumping}

\subsubsection{Temporal evolution of spin signals}
\begin{figure}[htb]
\includegraphics[height=4.8cm]{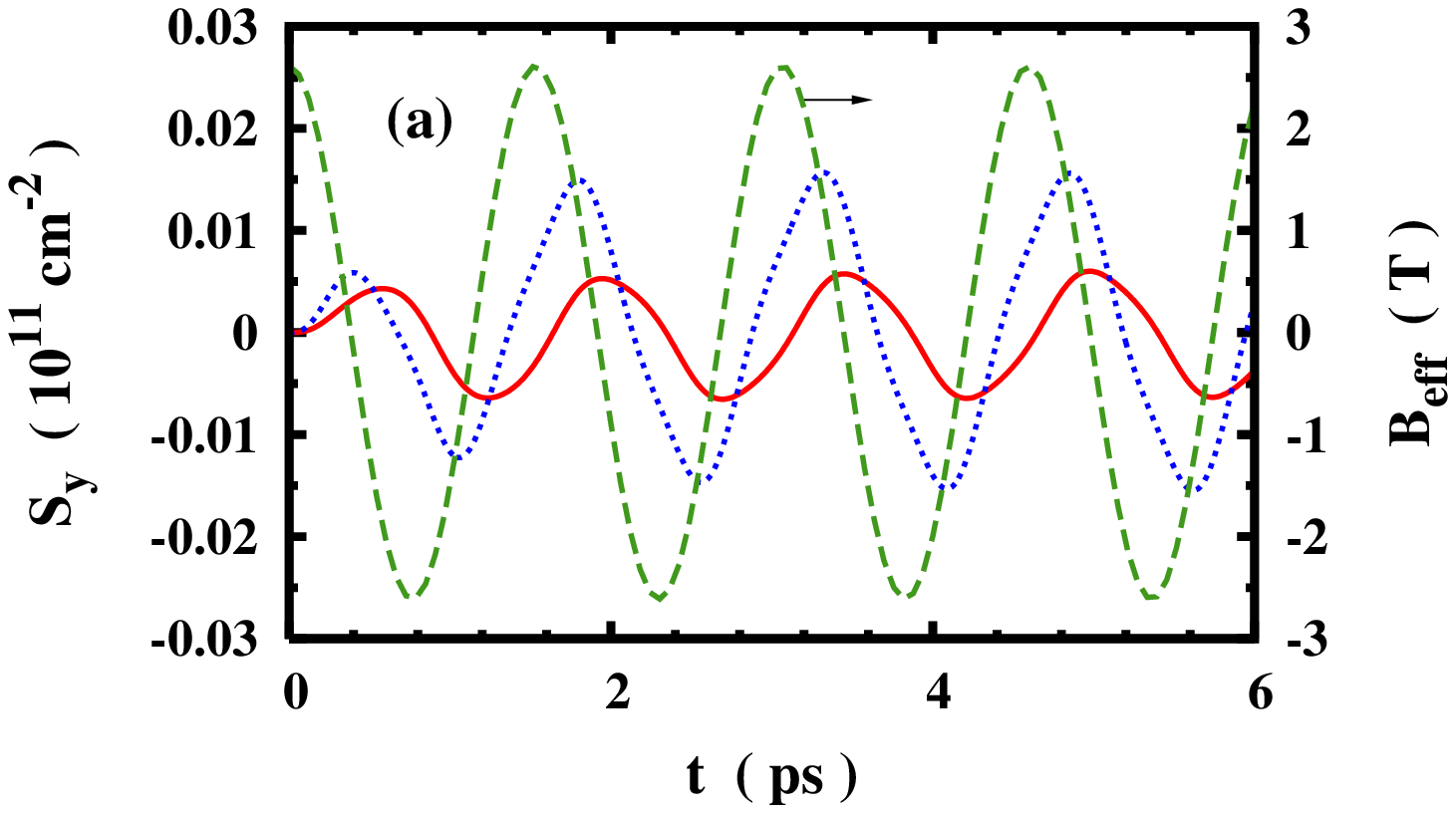}
\includegraphics[height=5.cm]{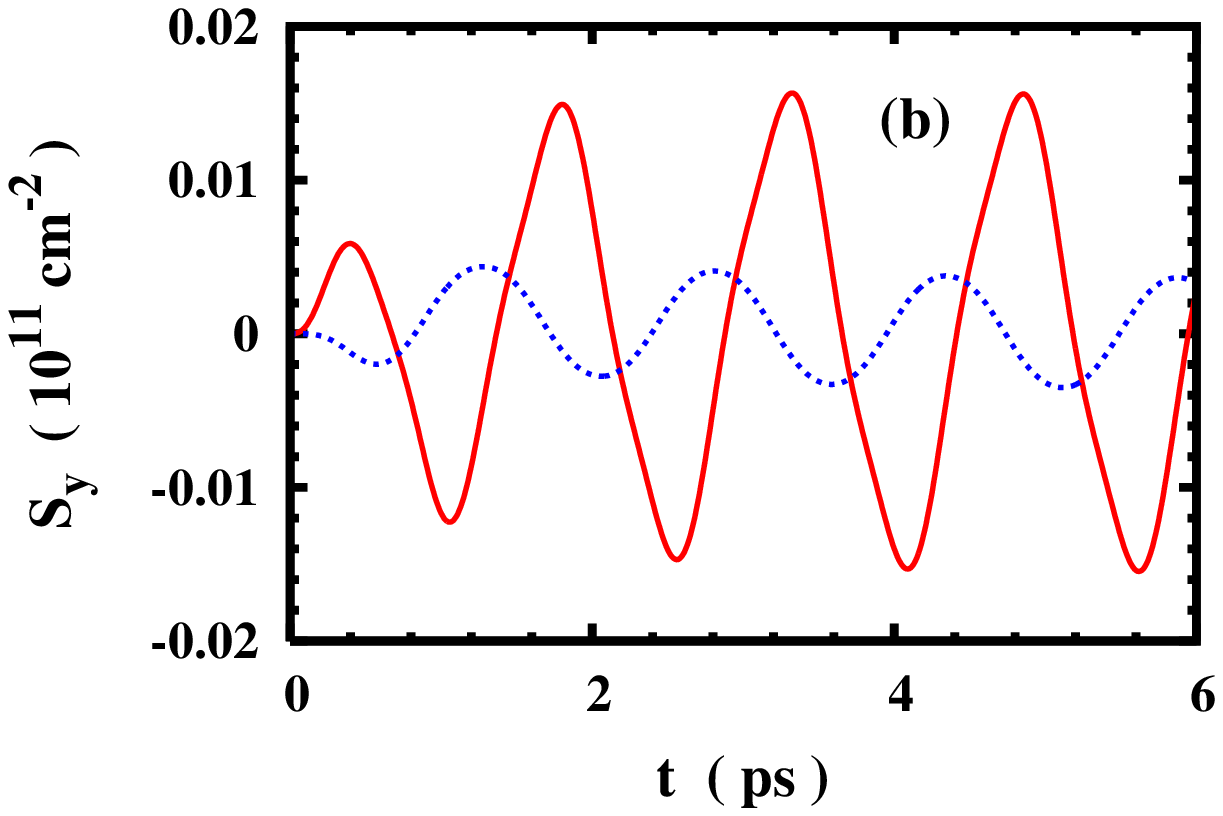}
\caption{ (Color online) (a) Spin polarization along $y$ axis,
$S_y$, as function of
  time with zero initial spin polarization for $E=0.5$\ kV/cm (solid curve) and
1.0\ kV/cm (dotted curve). $T=50$\ K and
$N_i=0.05 N_e$. The dashed curve is the
  THz-field-induced effective magnetic field $B_{\mbox{eff}}$ with $E=1.0$ kV/cm. Note
  that the scale of the dashed curve is on the right hand side of the frame.
(b) $S_y$ {\em vs.} time  for $E=1.0$\ kV/cm
  with  $B_{\mbox{eff}}$  included
  (solid curve) and excluded (dotted curve).}
\label{fig:Spinpump}
\end{figure}

In Fig.\ \ref{fig:Spinpump}(a), we plot the spin polarization along
the $y$ axis, $S_y$, as a
function of time when the initial spin polarization is zero for $E=0.5$
(solid curve) and 1.0 (dotted curve)\ kV/cm. The spin polarizations
along $z$- and $x$-axes are always zero. We also plot
the THz-field-induced effective magnetic field,
$B_{\mbox{eff}}$ [Eq.\ (\ref{beff})], as  dashed curve in the
figure. It is noted that the THz field pumps a large (several percent)
spin polarization which oscillates at the same frequency with the THz
field. This feature coincides with what predicted in the previous work
where no dissipations are considered.\cite{Cheng} Nevertheless,  it is
interesting to see that there is a delay of $S_y$ with respect to the
THz-field-induced effective magnetic field $B_{\mbox{eff}}$,
which is different from the
dissipation-free case. The time dependence of $S_y$ falls into the
general form
\begin{equation}
  S_y(t) = \sum_{n>0}S^{0n}_y\cos[n\Omega(t-t^n_d)],
\label{Sy}
\end{equation}
with $S^{0n}_{y}$ and $t^n_d$ denoting the amplitude and the delay time
respectively. The delay times are due to the retarded response of the
spin polarization to the spin pumping caused by the THz field.
This can be revealed by the following simplified
analysis. Approximately, $S_y$ satisfies the following equation,
\begin{equation}
  \partial_{t} S_y = - \left( S_y - \bar{S}_y(t) \right) / \tau_s ,
\label{Syt}
\end{equation}
where $\bar{S}_y(t)$ is the instantaneous equilibrium spin
polarization induced by $B_{\mbox{eff}}(t)$
due to Pauli spin paramagnetism. The factor $1/\tau_s$ represents
the spin relaxation. Under the initial condition $S_y(0)=0$, the
equation has the following solution
\begin{equation}
  S_y(t) = \int_{0}^{t} \!\!d{t^{\prime}} \frac{\bar{S}_y(t^{\prime})}{\tau_s}
  e^{-(t-t^{\prime})/\tau_s}.
\end{equation}
When the THz field is strong, the instantaneous equilibrium spin
polarization  has the form of multi-frequency dependence: $\bar{S}_y(t)
=\sum_n \bar{S}^n_y e^{in\Omega t}$, as demonstrated in Ref.\
\onlinecite{Cheng}. As the effective magnetic field is in the form of
cosine function, $\bar{S}_y(t)$ should be in the form
$\bar{S}_y(t) =\sum_{n>0} 2\bar{S}^n_y \cos(n\Omega t)$, where
$\bar{S}^n_y$ is real. The solution of $S_y(t)$ at $t\gg\tau_s$ is hence
given by
\begin{equation}
  S_y(t) = \sum_{n>0} \frac{2\bar{S}^n_y}{\sqrt{(n\Omega\tau_s)^2 + 1}}
  \cos[n\Omega(t-t_n)],
\label{SyF}
\end{equation}
with
\begin{equation}
t_n= \arctan(n\Omega\tau_s)/(n\Omega).
\label{tn}
\end{equation}
Comparing the above equation with Eq.\ (\ref{Sy}), one obtains
$S^{0n}_{y} = 2\bar{S}^n_y/\sqrt{(n\Omega\tau_s)^2
  +1}$ and $t^n_d=t_n$.
The delay time $t^n_d$ is indeed due to the retarded response
of the spin polarization to the spin pumping. In the limit of
$\tau_s\to 0$, one has  $S_y(t)=\bar{S}_y(t)$, i.e.,  the spin
polarization completely follows the spin pumping due to the THz field.
This is exactly the property of the results obtained in the previous
dissipation-free studies.\cite{Cheng,Jiang,yzhou}

Typically $n$ runs in the range of $n=1,2$ in the parameter regime of
our investigation. For small THz field strength, $E\lesssim0.4$ kV/cm, only the term with
$n=1$  contributes to $S_y$. For larger field strength, the term with $n\ge2$
also contributes and the peak of $S_y(t)$
is not symmetric
any more. Although the term with $n\ge2$ may also contribute, the most important
 contribution still comes from $n=1$ term. Consequently
  $S_y$ signal still has good periodic behavior.

It should be mentioned that under the RWA, the kinetic spin Bloch equations
in the Floquet picture is explicitly time-independent.\cite{Jiang2,FM}
Thus, the steady-state density operator in the
 Floquet picture becomes  time-independent\cite{FM} and
 the time-dependence of the spin
polarization $S_y$ in the steady state becomes totally determined by the
time-evolution of the Floquet states,
which completely follow the spin pumping due to the
THz field. As a result, the RWA loses the
important information of the retardation of the spin
polarization to the THz field. In our study, we go beyond the RWA.

As pointed out after Eq.\ (\ref{hamilton}) that the
anisotropic term $\gamma_E k_x\Omega\cos(\Omega t)$ in the Hamiltonian
breaks down $k_x\to -k_x$ symmetry. However,
without scattering, the density matrix in the Floquet picture does not
change with time and is determined solely by its initial value. For the
choice of isotropic initial distribution, the $k_x\to -k_x$ asymmetry
of the density matrix never shows up. In the presence of scattering, the density matrix should
show the asymmetry of the Hamiltonian. Consequently, the average of
$k_x$ of the electron system, $\langle k_x\rangle$, is nonzero.
Below we will show that, quite remarkably, the  scattering terms within the RWA
{\em keeps} the symmetry of $k_x\to -k_x$ and the scattering terms which
do not keep the symmetry only appear in the time-dependent (beyond RWA)
scattering terms.

Look at, e.g., the electron-impurity scattering [Eq.~(\ref{scatei})],
the weight of the $n$-th sideband-modulated
scattering is
$P_{n}^{(\eta\eta_1\eta_2\eta_3)}=S^{(\eta\eta_1)}_{{\bf k},{\bf
    k}^{\prime}}(t,0)S^{(n)(\eta_2\eta_3)}_{{\bf k}^{\prime},{\bf
    k}}$. As the main source of the sideband effect is  the
term $\gamma_E
k_x\Omega\cos(\Omega t)$  in the Hamiltonian [Eq.~(\ref{hamilton})], the
 weight is approximately
\begin{eqnarray}
P_{n}^{(\eta\eta_1\eta_2\eta_3)}&=&\delta_{\eta\eta_1}\delta_{\eta_2\eta_3}e^{i[(\varepsilon_{{\bf k}}-\varepsilon_{{\bf
      k}^{\prime}})t+\gamma_E\sin(\Omega t)(k_x-k_x^{\prime})]}
\nonumber \\
&&\mbox{}\times J_n[\gamma_E(k_x^{\prime}-k_x)].
\end{eqnarray}
$P_{n}$ can be further decomposed into the time-independent $P_{n}^{in}$ and the
time-dependent $P_{n}^{d}=P_{n}^{d1}+P_{n}^{d2}$ parts (omitting the superscripts of $\eta$):
\begin{equation}
  P_{n}^{in} = J_{n} \left(\gamma_E(k_x^{\prime}-k_x)\right)
  J_{n}\left(\gamma_E(k_x^{\prime}-k_x)\right),
\label{Pin}
\end{equation}
\begin{eqnarray}
  P_{n}^{d1} &=& \sum_{m:\mbox{even}\ m\ne 0}
  J_{n+m}\left[\gamma_E(k_x^{\prime}-k_x)\right] \nonumber \\
&&\mbox{} \times
  J_n\left[\gamma_E(k_x^{\prime}-k_x)\right] e^{-im\Omega t}, \\
  P_{n}^{d2} &=&  \sum_{m:\mbox{odd}}
  J_{n+m}\left[\gamma_E(k_x^{\prime}-k_x)\right]
\nonumber \\
&&\mbox{} \times
  J_n\left[\gamma_E(k_x^{\prime}-k_x)\right] e^{-im\Omega t}.
\label{Pt}
\end{eqnarray}
It is seen that under the transformation: $k_x^{\prime}\to
-k_x^{\prime}$ and $k_x\to -k_x$,  $P_{n}^{in}$ and  $P_{n}^{d1}$
are invariant but $P_{n}^{d2}$ is changed. This
indicates that a portion of the time-dependent (beyond RWA) scattering
terms break down the $k_x\to-k_x$ symmetry. With these scattering
terms, the density matrix should evolve to be asymmetric in $k_x$
direction. This leads to $\langle k_x\rangle\ne
0$.  $\langle k_x\rangle$ should also oscillate with
time as $P_{n}^{d2}$ does.

In the presence of the SOC, $\langle
k_x\rangle$ leads to a second effective magnetic field:
\begin{equation}
\label{beff2}
B_{av}(t)=2\alpha_R\langle
k_x\rangle/(|g|\mu_{B}).
\end{equation}
Indeed, we find that the spin polarization
$S_y$ is still nonzero when $B_{\mbox{eff}}$ is turned off by omitting
the corresponding term in the Hamiltonian. In Fig.\
\ref{fig:Spinpump}(b), we plot $S_y(t)$ for both cases with and
without $B_{\mbox{eff}}$. It is seen that $S_y$  is
nonzero when $B_{\mbox{eff}}$ is excluded, although the
amplitude is reduced. This spin polarization is induced by $B_{av}$
via Pauli paramagnetism. The results indicate that $B_{av}$
oscillates with time and is smaller than $B_{\mbox{eff}}$.
Moreover, there is a change in the delay of the oscillation due to
different time-dependence of $B_{av}$ compared to $B_{\mbox{eff}}$.
This difference also contributes to the delay of the
spin polarization which is induced by the {\em total} effective
  magnetic field ${\cal B}$ (${\cal B}\equiv B_{\mbox{eff}}+B_{av}$).

Finally, it is found that a small initial spin polarization ($\sim
4\%$) along the $z$ axis makes marginal effect on the time dependence
of $S_y$. %This confirms %the phenomenological Eq.\ (\ref{Syt}).

\subsubsection{Steady-state spin polarization}
In this subsection, we discuss the dependence of the amplitude of the
steady-state spin polarization $S_y^0$ (the peak value of
$S_y$) on the  THz field.

\begin{figure}[htb]
\includegraphics[height=6.5cm]{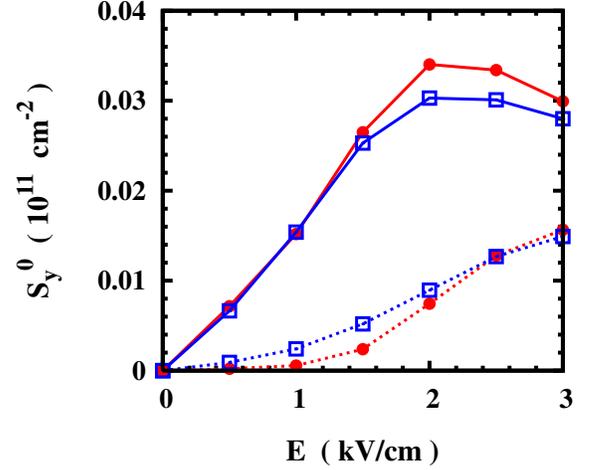}
\caption{(Color online) Dependence of the ASSSP
 on THz field strength for $T=50$\ K (solid
  curve with $\bullet$) and 100\ K (solid curve with $\square$) without
 impurities  ($N_i=0$). The dotted curves are % with $\bullet$  ($\square$) is
the same as the solid ones % with $\bullet$  ($\square$),
but without the THz-field-induced effective
  magnetic field $B_{\mbox{eff}}$.}
\label{fig:Sy_diffKdiffNi}
\end{figure}

In Fig.\ \ref{fig:Sy_diffKdiffNi} we plot the amplitude of the steady-state spin
polarization (ASSSP) as a function of THz field strength for
the cases with and without the  THz-field-induced effective
  magnetic field $B_{\mbox{eff}}$. Two typical
lattice temperatures $T=50$\ K and $100$\ K are
investigated with the
impurity density $N_i=0$. It is seen that the ASSSP first increases
then decreases with the strength of the THz field.  The effective
magnetic field increases with the THz field strength. According to
Pauli paramagnetism, however, the spin polarization should always
increase with the magnetic field. Here the decrease of the ASSSP
mainly originates from the hot-electron effect. To elucidate this
point, we plot the hot-electron temperature $T_e$ in Fig.\
\ref{fig:Te1_diffKdiffNi} (the method used to obtain $T_e$ is given
in Appendix\ C). It is seen that the hot-electron temperature
increases with the THz field strength. The increase of the
hot-electron temperature decreases the induced spin polarization
according to Pauli paramagnetism. It is noted from the figure that
the largest ASSSP can be $\sim3.5\times 10^{9}$\ cm$^{-2}$ which
corresponds to a large spin polarization of $7$\ \%. This indicates
that the intense THz field is a very efficient tool in generating
spin polarization. It can be noticed in Fig.\ \ref{fig:Sy_diffKdiffNi}
that the ASSSP is smaller at higher temperature. The decrease of the
ASSSP is due to the increase of the hot-electron temperature with
the lattice temperature, as indicated in Fig.\
\ref{fig:Te1_diffKdiffNi}.

\begin{figure}[htb]
\includegraphics[height=6.5cm]{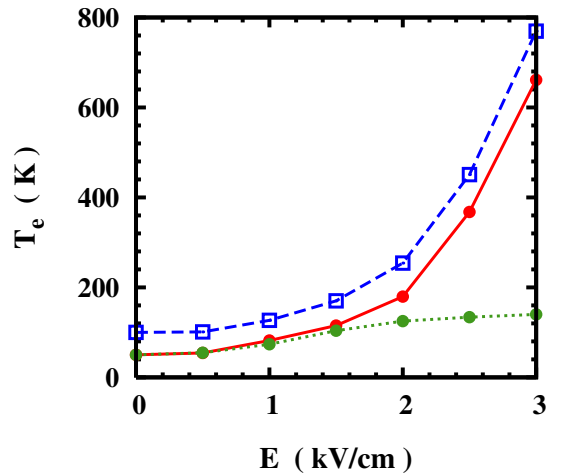}
\caption{(Color online) Hot-electron temperature $T_e$ as function
 of THz field strength for $T=50$\ K (solid curve with $\bullet$) and
100\ K (dashed curve with $\square$) without  impurities
($N_i=0$). The dotted curve with $\bullet$ is the same as the solid
one, but with  only  $n=0,\pm 1$ allowed in the
sideband-modulated scattering.}
\label{fig:Te1_diffKdiffNi}
\end{figure}

It should be pointed out that, differing from our previous study on spin
relaxation in quantum dots,\cite{Jiang2} here the $n$-th
sideband-modulated scattering rate differs little from each other. The energy
conservation (the $\delta$ functions in the scattering terms) gives
different final state for different $n$ with given initial state, thus
the momentum  transferred  into the system
can change effectively with $n$. However, the matrix
elements of all the scattering mechanisms vary slowly with the
momentum  due to the screening and the quantum confinement along the
growth direction. Consequently, the sideband-modulated scattering
rate varies slowly with $n$ and the manipulation of
the spin relaxation via
 sideband modulation of the spin-flip scattering does not
apply in 2DES.  In 2DES, the main effect of the
  sideband-modulated scattering is the hot-electron effect. As we
have pointed out, the $n$-th sideband-modulated scattering tends to make the distribution
be flatter in the energy range of $n\Omega$, which thus leads to
the hot-electron effect. In Fig.\ \ref{fig:Te1_diffKdiffNi}, we also plot the hot-electron
temperature when the summations of $n$ in the scattering terms [Eqs.\
  (\ref{scatei}), (\ref{scatep}) and (\ref{scatee})] are restricted to
$n=0,\pm 1$. In the previous studies on the effect of THz field on
spin dynamics, only these processes are considered, where the THz
field is weak.\cite{Ganichev1,Ganichev2,Ganichev3} It is
seen that the hot-electron temperature is largely reduced by
the number of sideband involved in the scattering, especially
when the THz field is strong and hence the sideband-modulated scattering
with $|n|>1$ is important.
This confirms the important role of sideband-modulated
scattering to the hot-electron effect.

\begin{figure}[htb]
\includegraphics[height=6.5cm]{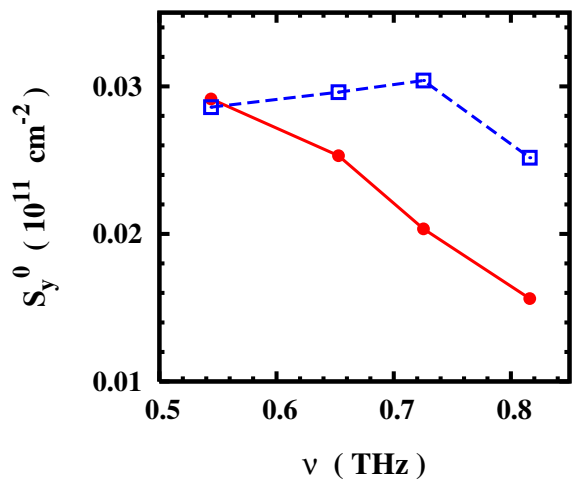}
\caption{(Color online) Dependence of the ASSSP
 on THz field frequency for $E=1.5$\ kV/cm
  ($\bullet$) and $2.5$\ kV/cm ($\square$). $T=100$\ K and
  $N_i=0$. }
\label{fig:Sy_50Kdiffomega}
\end{figure}

The dotted curves  in Fig.\ \ref{fig:Sy_diffKdiffNi}, is
the ASSSP calculated without the THz-field-induced effective magnetic
field $B_{\mbox{\tiny eff}}$. As analyzed before, here the ASSSP is induced by
$B_{av}$. The contribution of $B_{av}$ becomes more important when THz
field strength increases. This is because that the terms $P_{n}^{d2}$ increase with the
sideband effect which increases with THz field
strength. Moreover, it is seen that the ASSSP due to $B_{av}$ is
larger at higher temperature (100\ K in the figure) when the THz field
strength is small. This is because the scattering terms leading to
the breakdown of $k_x\to-k_x$ symmetry increase as the electron--LO-phonon
scattering is more efficient at $T=100$\ K. However, when the THz field
strength is larger, the hot-electron effect becomes more important. (We find that
the hot-electron temperature changes little when $B_{\mbox{eff}}$ is
removed.) The hot-electron effect also reduces the ASSSP induced
by $B_{av}$. At large THz field strength this effect becomes more
important and the difference of  ASSSPs at 50 and $100$\ K becomes marginal.

\begin{figure}[htb]
\includegraphics[height=6.5cm]{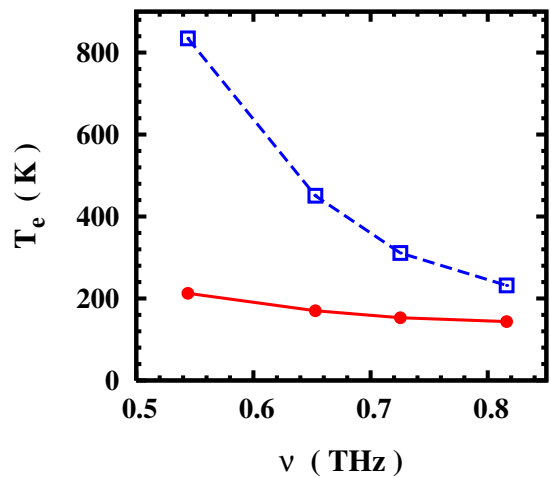}
\caption{(Color online) Hot-electron temperature $T_e$ as function of the
  THz frequency for $E=1.5$\ kV/cm
  ($\bullet$) and $2.5$\ kV/cm ($\square$). $T=100$\ K and
  $N_i=0$.}
\label{fig:Te1_50Kdiffomega}
\end{figure}

We now turn to investigate the dependence of the ASSSP on THz frequency. It is
noted that the amplitude of the THz-field-induced effective
magnetic field $B_{\mbox{eff}}$ [Eq.\ (\ref{beff})] decreases with
THz frequency. However, the hot-electron effect induced by the sideband effect
which increases with $\gamma_E\Omega=eE/(m^{\ast}\Omega)$,
also decreases with THz frequency. These two effects again compete with each other. In Fig.\
\ref{fig:Sy_50Kdiffomega} we plot the ASSSP as function of the THz
frequency for two cases: $E=1.5$ and $2.5$\ kV/cm with $T=100$\ K.
It is seen that for the case with $E=2.5$\ kV/cm, the ASSSP first increases
then decreases with the THz frequency due to the competition
of the two effects. To examine the hot-electron effect, we also plot
the hot-electron temperature in Fig.\ \ref{fig:Te1_50Kdiffomega}. It
is seen that the hot-electron temperature decreases with the THz
frequency. For large THz frequency, the hot-electron effect is
marginal and thus the ASSSP decreases with the THz frequency as the
THz field induced effective magnetic field does. For smaller THz
frequency the hot-electron effect becomes dominant and the ASSSP
increases with the THz frequency. Consequently there is a peak frequency
where the ASSSP reaches the maximum. For the case with $E=1.5$\ kV/cm, there should
be a peak with the peak frequency being much smaller than the frequency we calculated.
This is because the hot-electron effect
is much weaker than the case with $E=2.5$\ kV/cm (see Fig.\
\ref{fig:Te1_50Kdiffomega}).

\subsection{Spin dynamics with finite initial spin polarization}

\subsubsection{Temporal evolution of the spin signals}

\begin{figure}[htb]
\includegraphics[height=3.6cm]{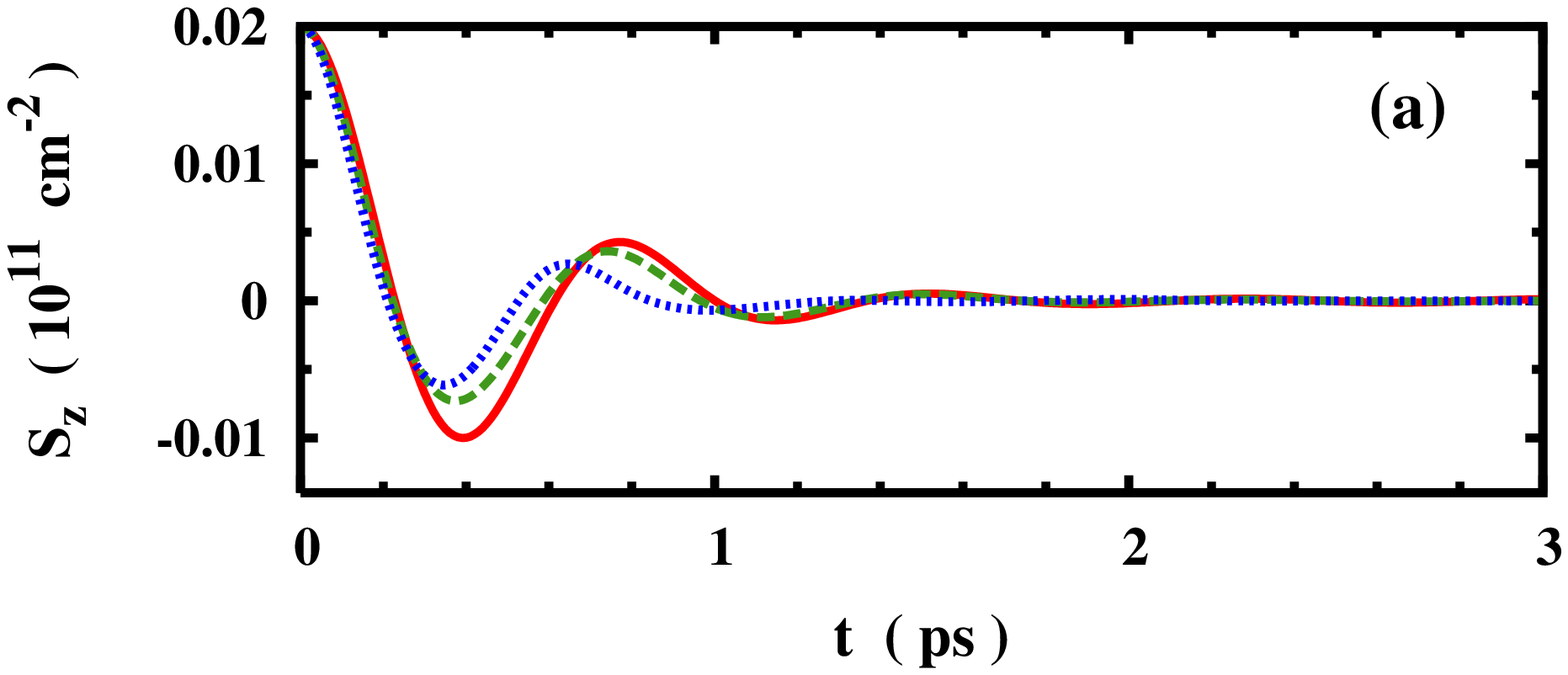}
\includegraphics[height=3.6cm]{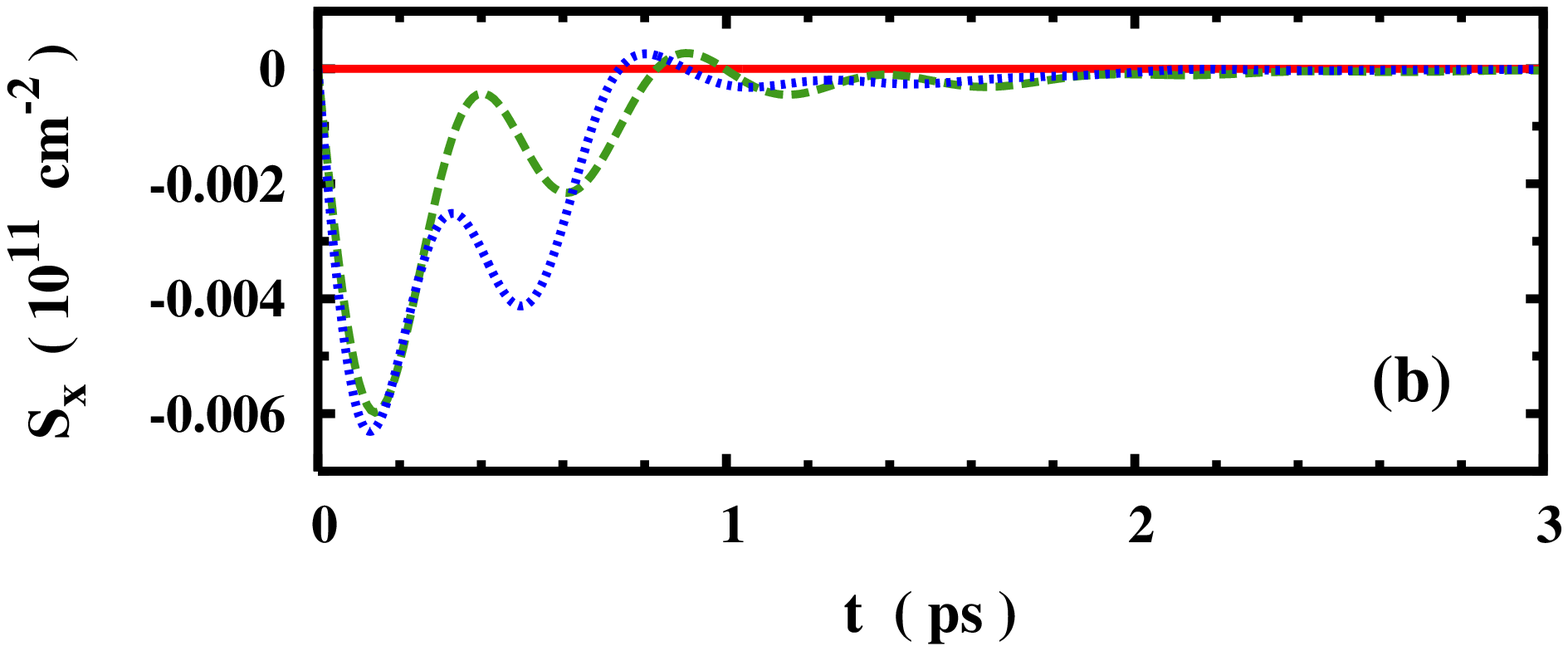}
\includegraphics[height=3.6cm]{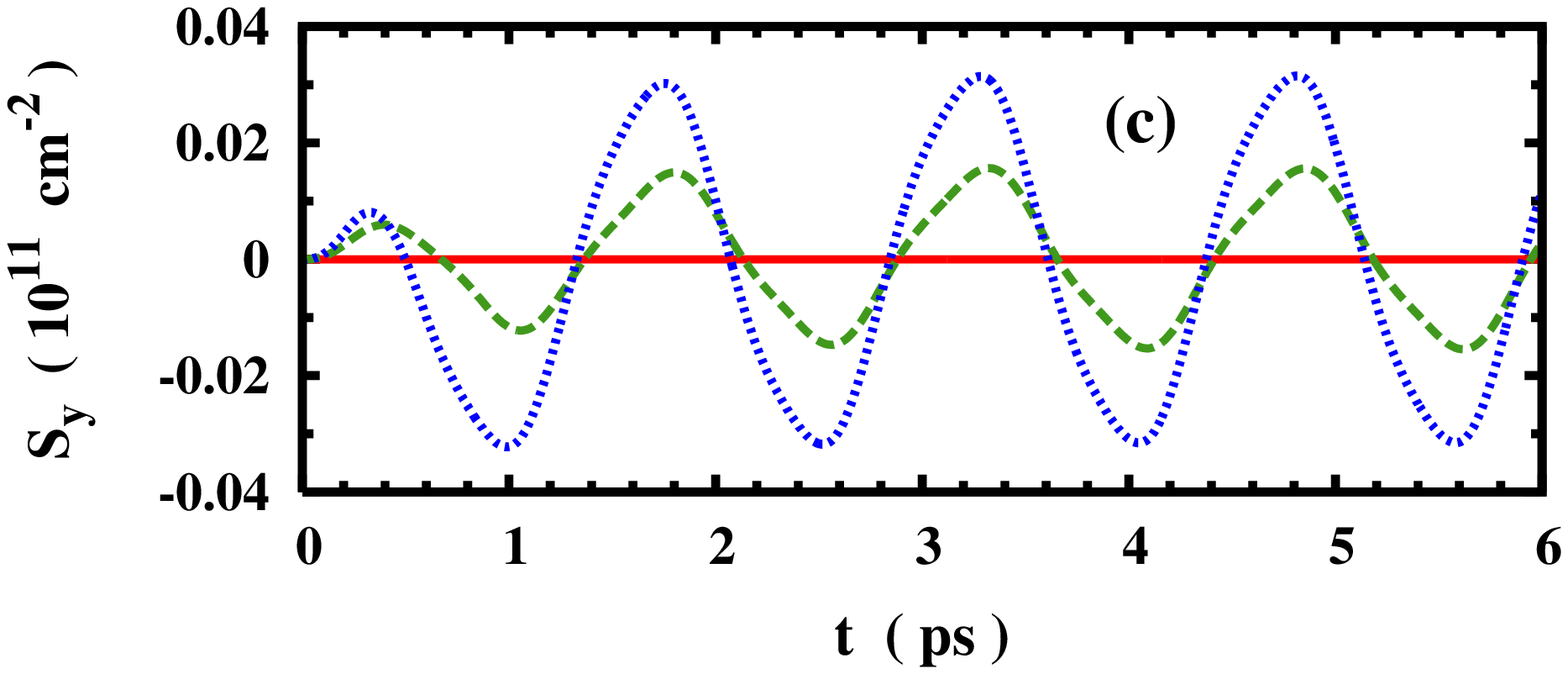}
\caption{(Color online) Temporal evolution of the spin signals (a):
  $S_z$; (b): $S_x$; and (c): $S_y$ for THz field strengthes: 0\ kV/cm
  (solid curves), 1\ kV/cm (dashed curves), and 2\ kV/cm (dotted
  curves). $N_i=0.05 N_e$ and $T=50$\ K.}
\label{fig:Sxyz_t}
\end{figure}

In Fig.\ \ref{fig:Sxyz_t}, we plot the temporal evolutions of the spin signals
 along $z$, $x$ and $y$ axis at
different THz field strengths. The initial spin polarization
is taken to be 4\ \% along the $z$ axis. In Fig.\ \ref{fig:Sxyz_t}(a),
 one finds that $S_z$ exhibits
oscillatory decay. This resembles the low temperature spin decay
observed in Refs.\ \onlinecite{wu-Exp-HighP} and \onlinecite{Harley},
which is due to the large spin-orbit effective magnetic field and weak
scattering,\cite{wu-hole} i.e., the system is
in the weak scattering limit.
It is noted that $S_z$ decays faster when
the THz field strength increases. %first to 1 kV/cm (dashed curve) and then to 2 kV/cm (dotted curve).
Moreover, the spin oscillation frequency
also increases as indicated by the left-shift of the peak around 0.7\ ps,
which is due to the total effective
 magnetic field ${\cal B}$ induced
by the THz field. From Fig.\ \ref{fig:Sxyz_t}(b), one notices  that
a small value of $S_x$ is excited but eventually decays to zero. This
is again due to the effective magnetic field ${\cal B}$,
which rotates $S_z$ to $S_x$. The first peak value of $S_x$ increases with
the THz field strength. Without THz field, $S_x\equiv 0$. In Fig.~\ref{fig:Sxyz_t}(c), it is
seen that $S_y$ is also induced and reaches a non-vanishing
oscillatory value after $\sim 3$\ ps evolution, similar to what
observed in Fig.\ \ref{fig:Spinpump} where the initial spin polarization is zero.

\subsubsection{SRT}

\begin{figure}[htb]
\includegraphics[height=6.5cm]{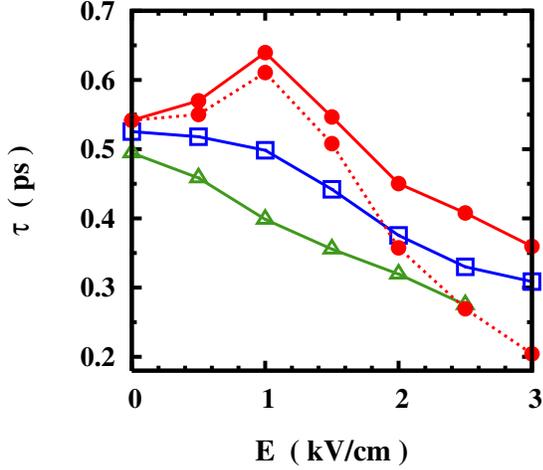}
\caption{(Color online) Dependence of the SRT $\tau$
  on THz field strength for impurity densities: $N_i=0$ ($\bullet$);
  $N_i=0.02N_e$ ($\square$); $N_i=0.05N_e$ ($\triangle$). Solid
  curves: from full calculation; Dotted curve: from the calculation
 without the THz-field-induced
  effective magnetic field $B_{\mbox{eff}}$.}
\label{fig:SRT_50KdiffNi}
\end{figure}

In Fig.\ \ref{fig:SRT_50KdiffNi}, the SRT,
which is extracted via fitting  the
exponential decay of the envelop of $S_z$, is plotted as
function of THz field strength for different impurity densities
$N_i=0$, $0.02N_e$, and $0.05N_e$, with lattice temperature $T=50$\ K.

\begin{figure}[htb]
\includegraphics[height=6.5cm]{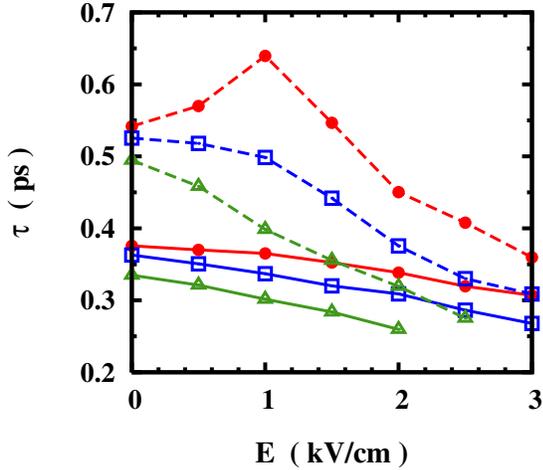}
\caption{(Color online) Dependence of the SRT $\tau$
  on THz field strength for impurity densities: $N_i=0$ ($\bullet$);
$N_i=0.02N_e$ ($\square$); and $N_i=0.05N_e$ ($\triangle$). $T=100$\ K
  (solid curves) and $50$\ K (dashed curves).}
\label{fig:SRT_100KdiffNi}
\end{figure}

We first discuss the case with $N_i$=0 (solid curve with
$\bullet$). It is noted that the SRT first increases then decreases
with the THz field strength. The underlying physics is that there are two
consequences of the THz field: (i) the total THz field-induced
effective magnetic field ${\cal B}$; (ii) the
hot-electron effect. Effect (i) can
give a magnetic field as large as several Tesla (2.6\ T per
1~kV/cm THz field with $\nu=0.65$~THz,
of which the corresponding Zeeman splitting is as
large as 2.2~meV). This effective magnetic field blocks the
inhomogeneous broadening from the Rashba SOC.
It thus elongates the
SRT.\cite{opt-or,wu-later} The main consequences of effect (ii)
are the enhancement of momentum scattering as well as the
inhomogeneous broadening as the electrons distribute
 on larger ${\bf k}$ states where the SOC is larger.
Enhancement of the inhomogeneous broadening
 shortens the SRT according to our
previous studies.\cite{wu-later,wu-hole,wu-lowT}
It is found that in the strong scattering limit, 
SRT increases with the momentum scattering.\cite{opt-or,wu-review}
However, it is demonstrated in Ref.\ \onlinecite{wu-hole} that in
the weak or intermediate scattering limit, the SRT
 decreases with increasing the momentum scattering. In
our case, due to the large Rashba SOC parameter,
the system is in the weak/intermediate scattering regime.
This can be further checked by the fact that the SRT decreases
when the electron-impurity scattering is strengthened by increasing
the impurity density, as  shown in Figs.\ \ref{fig:SRT_50KdiffNi}
and \ref{fig:SRT_100KdiffNi}. Thus, effect (ii) shortens the SRT. The
two effects compete with each other and hence the SRT varies {\em
  non-monotonically} with the THz field strength: For small THz
field strength the increase of the total  THz field-induced effective
magnetic field ${\cal B}$ is dominant. As a result,
the SRT increases; For large field strength, the hot-electron
effect becomes more important. Consequently
the SRT decreases.

To further elucidate the influence of effect (i),
 we remove part of the THz field-induced effective magnetic
field, $B_{\mbox{eff}}$, by excluding the term
$\alpha_R\hat{\sigma}_yeE\cos(\Omega t)/\Omega$ from the
Hamiltonian, and then calculate the
SRT. We plot the obtained SRT as dotted curve in Fig.\
\ref{fig:SRT_50KdiffNi}. It is seen that the SRT is reduced, especially
at large  THz field strength. It is checked that the
hot-electron effect changes little when $B_{\mbox{eff}}$ is removed as it
is not the main source of the hot-electron  effect. The
results confirm that the THz field-induced effective magnetic field indeed
increases the SRT.

\begin{figure}[htb]
\includegraphics[height=6.5cm]{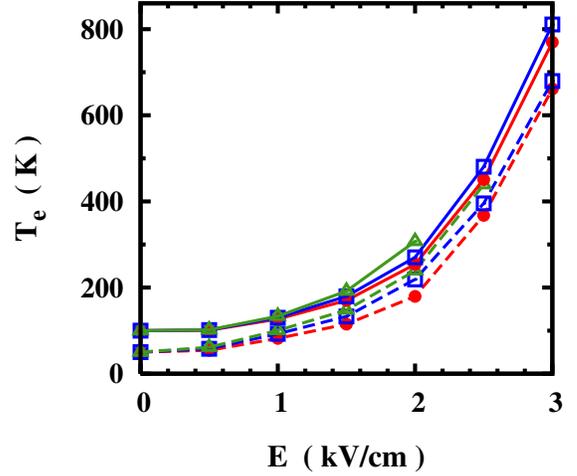}
\caption{(Color online) Dependence of the hot-electron temperature $T_e$
  on THz field strength for different impurity densities: $N_i=0$ ($\bullet$
  ); $N_i=0.02N_e$ ($\square$); $N_i=0.05N_e$ ($\triangle$). $T=100$ K
  (solid curves) and $T=50$ K (dashed curves).}
\label{fig:Te_100KdiffNi}
\end{figure}

For the cases with $N_i=0.02N_e$ and $0.05N_e$, the
SRT decreases with the THz field strength monotonically. It is noted in Fig.\ 9
(solid curves) that the hot-electron temperature is larger at higher
impurity density. The enhancement of the hot-electron effect overcomes
the increase of the effect of the THz field-induced effective magnetic
field in these two cases, which leads to the monotonic decrease of the
SRT.

\begin{figure}[htb]
\includegraphics[height=6.5cm]{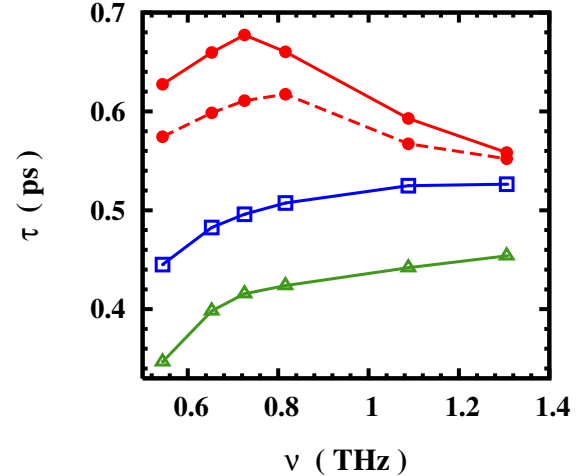}
\caption{(Color online) Dependence of the SRT $\tau$
  on THz frequency for impurity densities: $N_i=0$
  ($\bullet$); $N_i=0.02N_e$ ($\square$); and $N_i=0.05N_e$
  ($\triangle$). Solid curves: from full calculation; Dotted curve:
  from the calculation without the THz-field-induced effective
  magnetic field $B_{\mbox{eff}}$. $E=1$\ kV/cm and $T=50$\ K.}
\label{fig:SRT_50Kdiffomega}
\end{figure}

It should be mentioned that in the case of static
electric field, the hot-electron effect is more important at smaller
impurity density under a given electric
field.\cite{Lei1,wu-hot-e,wu-lowT} However, under intense THz field
the hot-electron effect is more pronounced at larger impurity density
where the sideband-modulated scattering, which can transfer the THz
photon energy into electron system, is stronger. Similar effects have
been reported by Lei in the study of charge transport under intense THz
field.\cite{Lei2}

\begin{figure}[htb]
\includegraphics[height=6.5cm]{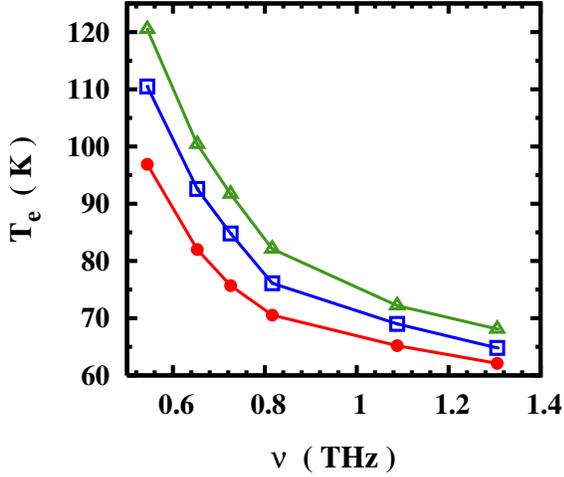}
\caption{(Color online) Dependence of the hot-electron temperature $T_e$
  on THz frequency for impurity densities: $N_i=0$
  ($\bullet$); $N_i=0.02N_e$ ($\square$); and $N_i=0.05N_e$
  ($\triangle$). $E=1$\ kV/cm and $T=50$\ K.}
\label{fig:Te_50Kdiffomega}
\end{figure}

We further discuss the  temperature dependence of the SRT.
In Fig.\ \ref{fig:SRT_100KdiffNi} we plot the SRT as function of
THz field strength at $T=100$\ K and
$50$\ K. One can see that the SRT at high temperature ($T=100$\ K)
is much smaller than that at low temperature  ($T=50$\ K).
 This is because the
electron--LO-phonon scattering at high temperature is much more efficient
than that at low temperature. The increase of scattering thus enhances
the hot-electron effect (see Fig.\ 9) and the enhancement of the
hot-electron effect reduces the SRT. It is also noted that for all three
impurity densities at $100$\ K, the SRTs decrease with the THz field
strength monotonically. This indicates that the hot-electron effect of
the THz field is dominant as the momentum scattering is strong.

We now turn to the THz frequency dependence of the SRT. As has been
demonstrated before, both the THz field-induced effective magnetic
field and the hot-electron effect decrease with the increase of
 the THz frequency and
they compete with each other on spin relaxation. In Fig.\
\ref{fig:SRT_50Kdiffomega}, we plot the SRT as
function of the THz frequency for three different impurity densities: $N_i=0$, $0.02N_e$, and $0.05N_e$. The lattice
temperature is $T=50$\ K. It is noted that for the impurity-free case,
the SRT first increases then decreases with the THz frequency, which
is similar to the dependence of the SRT on the THz field strength. This
is again due to the competition between the hot-electron effect and the THz
field-induced effective magnetic field. Similarly, for the cases
with $N_i=0.02N_e$ and $0.05N_e$, the SRT increases monotonically with the THz
frequency. When $B_{\mbox{eff}}$ is removed, the SRT for impurity-free
 case is reduced and the peak frequency where the SRT gets maximum  becomes larger.
This indicates the weakening of the THz field-induced effective magnetic field  since the hot-electron
effect changes little.

\begin{figure}[htb]
\includegraphics[height=6.5cm]{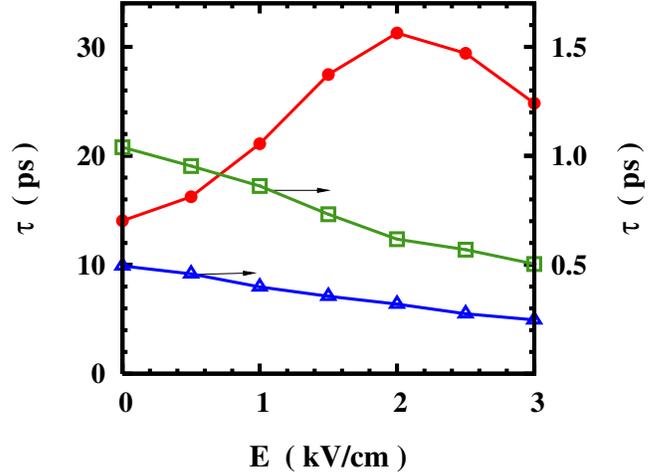}
\caption{(Color online) Dependence of the SRT $\tau$
  on THz field strength for different SOC parameters: $\alpha_R=1$\
  meV$\cdot$nm ($\bullet$); $\alpha_R=10$\ meV$\cdot$nm ($\square$);
and  $\alpha_R=30$\ meV$\cdot$nm ($\triangle$). $T=50$\ K and
  $N_i=0.05N_e$. Note that the scale for the curves with $\square$ and
$\triangle$ are on the right hand side of the frame.}
\label{fig:SRT_100Kdiffalpha}
\end{figure}

Finally, we discuss the dependence of the SRT on the Rashba SOC
parameter. In Fig.\ \ref{fig:SRT_100Kdiffalpha} we plot the SRT as
function of THz field strength at different Rashba parameters,
$\alpha_R=1$, 10, and 30\ meV$\cdot$nm.
The lattice temperature is taken to be $T=50$\ K and the impurity density is
$N_i=0.05 N_e$.   It is seen that for small Rashba SOC coefficient
$\alpha_R=1$\ meV$\cdot$nm, the SRT first increases then decreases with the THz
field strength. However, for large SOC,  the SRT decreases
monotonically with the THz field strength. As has been revealed
previously that in the presence of the impurity density $N_i=0.05N_e$, the hot-electron effect
dominates the SRT. The hot-electron effect leads to
the increase of both the scattering and the inhomogeneous
broadening. For the case with $\alpha_R=1$\ meV$\cdot$nm, which is in
the strong scattering regime, increase of scattering leads to
longer SRT; whereas the increase of inhomogeneous broadening leads to
shorter SRT. Therefore, the two effects compete with each other: the
SRT first increases due to the enhancement of scattering and
then decreases due to the increase of inhomogeneous broadening. This
behavior is similar to the case under a strong static electric
field.\cite{wu-hot-e} For the cases with larger SOC, which is in
the intermediate scattering regime, both  effects decrease the SRT.

\begin{figure}[htb]
\includegraphics[height=4.cm]{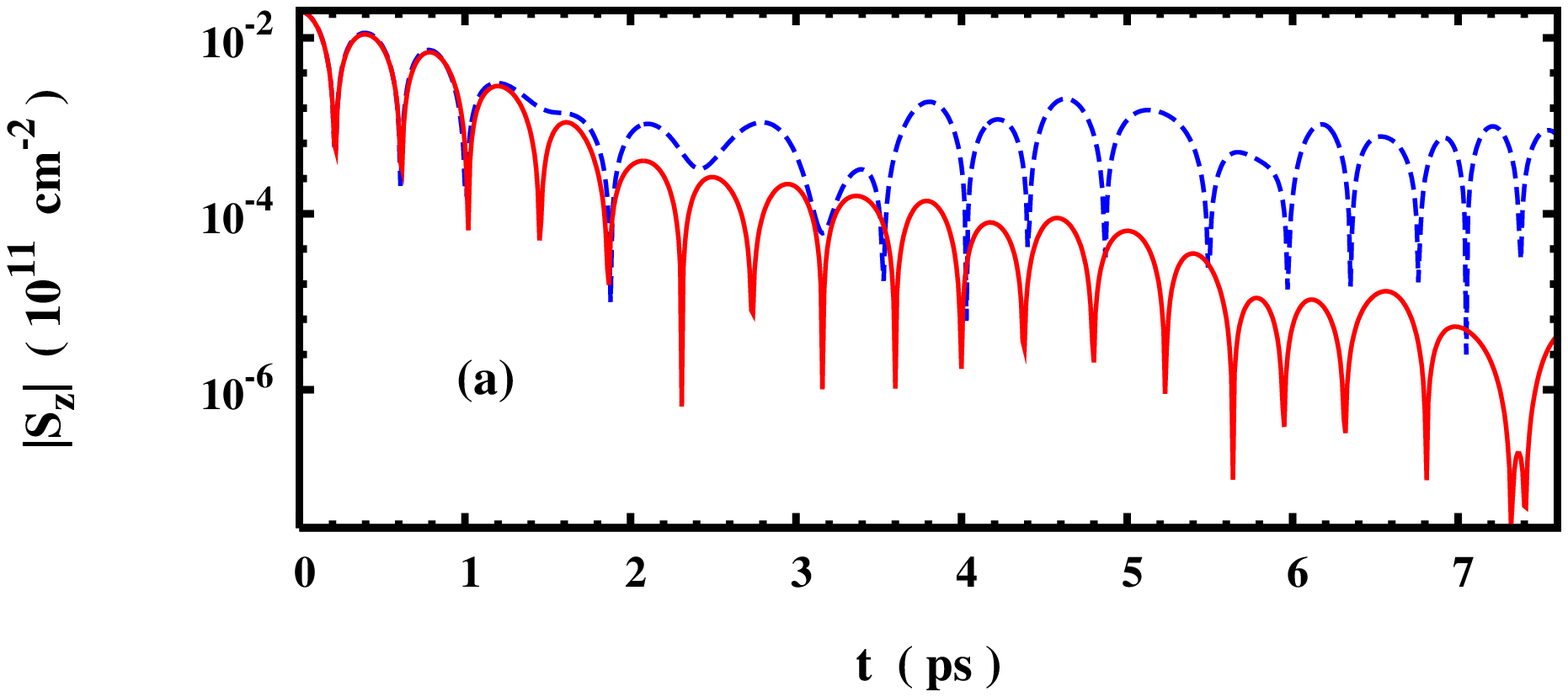}
\includegraphics[height=4.cm]{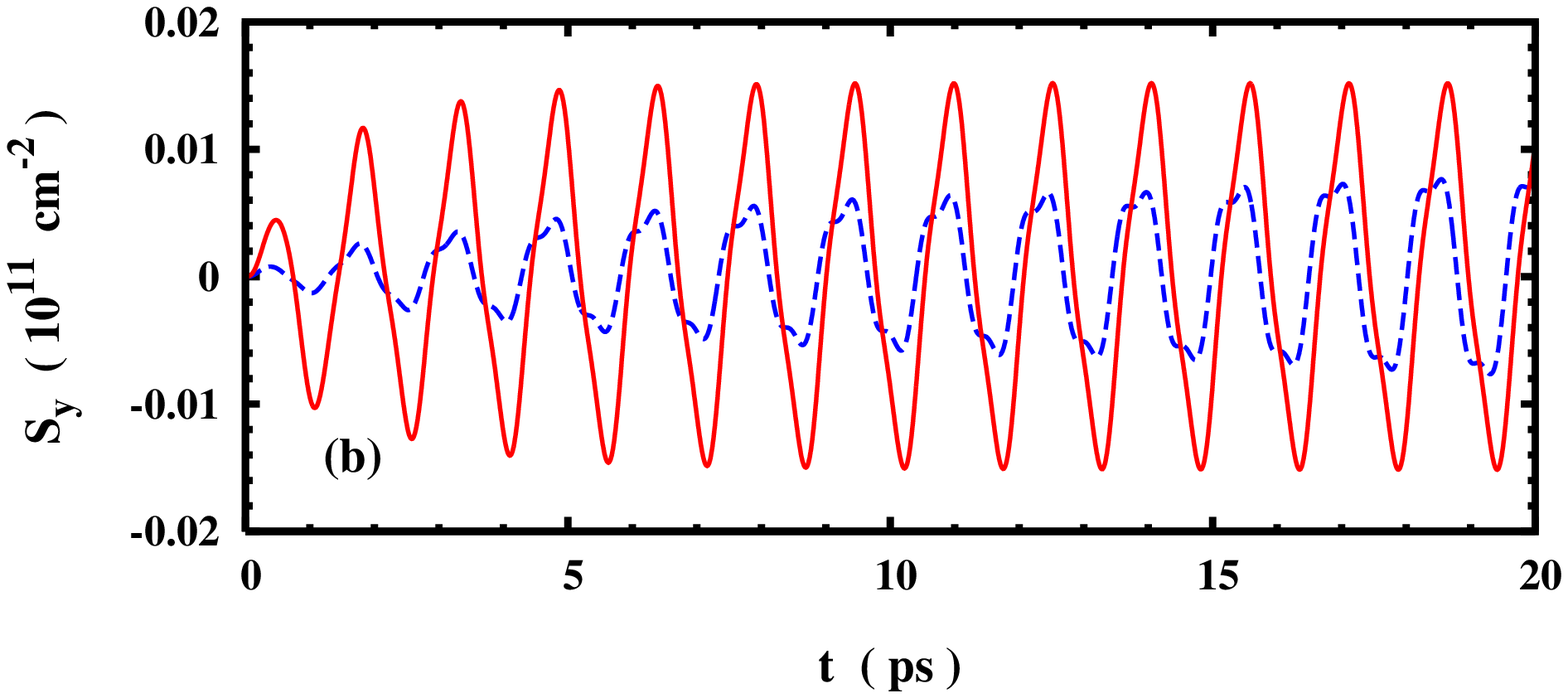}
\caption{(Color online) Temporal evolution of the spin signals
  calculated with the electron--electron scattering is included
  (solid curve) and excluded (dashed curve). (a):
  $|S_z|$; (b): $S_y$. $E=1$~kV/cm, $T=50$~K and $N_i=0$.}
\label{fig:Syz_t_Coulomb}
\end{figure}

\subsection{Effect of electron-electron Coulomb scattering}

Previously it has been shown that the electron-electron  Coulomb scattering
plays an important role in spin relaxation due to the DP
mechanism.\cite{wu-early,wu-hot-e,wu-helix,wu-lowT,wu-hole,Ivchenko,Ji}
Furthermore, for systems under strong electric field, the electron-electron
Coulomb scattering is
crucial for the electron system to establish its quasi-equilibrium
state.\cite{wu-hot-e} Here we demonstrate that it also has
nontrivial effects on the spin dynamics under intense THz field. In
Fig.\ \ref{fig:Syz_t_Coulomb}, we plot the temporal evolution of the
spin signals, $|S_z|$ and $S_y$, calculated with the
electron-electron scattering included (solid curve) and excluded
(dashed curve) under same initial distributions and conditions. It is
seen from Fig.~\ref{fig:Syz_t_Coulomb}(a) that in
the case with the electron-electron Coulomb scattering,  $|S_z|$ exhibits
good exponential decay, superimposed by the THz oscillations. Otherwise,
 the decay is non-exponential and the decay rate becomes much slower,
which indicates that the spin relaxation is markedly reduced.
It is further seen from  Fig.\ \ref{fig:Syz_t_Coulomb}(b) that
with electron-electron Coulomb scattering,  $S_y$
reaches the steady state  much faster with a larger
peak value. All these demonstrate the importance of
 the electron-electron Coulomb scattering to the spin dynamics.

\section{Conclusion and discussion}
\subsection{Conclusion}

In conclusion, we have developed the kinetic spin Bloch equations for 2DES
with Rashba SOC under intense THz laser fields,
with all the relevant scattering mechanisms such as the electron-impurity,
electron-phonon and electron-electron Coulomb scattering explicitly included. The
formalism is very general and can be applied to study  spin
kinetics in  many-body electron or hole system under strong
time-periodic driving fields with arbitrary SOC. Moreover,
 our formalism goes beyond the
 RWA treatment of the scattering.
By solving the kinetic spin Bloch equations
numerically, we investigate the effect of the intense THz fields on
the spin kinetics. We focus on the THz field-induced steady-state spin
polarization and the effect of the THz field on spin relaxation.

We first study the temporal evolution of the spin polarization under
intense THz field at zero initial spin polarization. We find that the
THz field can pump a {\em finite} steady-state THz spin polarization in
the presence of all relevant scattering. The spin polarization is
induced by the THz field-induced effective magnetic field in the
presence of SOC.
The maximum spin polarization in the steady state can be as large as 7\ \%,
which shows that the intense THz field is a very efficient tool in
generating spin polarization.

As our approach goes beyond the RWA treatment of the
scattering, we find some interesting features which are absent
in the RWA treatment. The first feature is that
there is always a retardation of the spin polarization
in response to the THz field-induced effective magnetic field.
Another feature is that, as the Hamiltonian breaks the $k_x\to
-k_x$ symmetry via the term $\gamma_E k_x\Omega\cos(\Omega t)$,
the average of $k_x$ over the electron system $\langle k_x\rangle$
becomes nonzero and oscillates with time. In the presence of SOC,
 $\langle k_x\rangle$ leads to another effective
magnetic field which also induces spin polarization. We find that,
remarkably, under the RWA,
the $k_x\to-k_x$ symmetry is still kept.

We further study the dependence of
the amplitude of the steady-state spin polarization on the THz field
for different lattice temperatures, impurity
densities and Rashba SOC parameters.
It is found that the main consequences of the THz field are: (i)
the hot-electron effect due to sideband-modulated scattering and
(ii) the THz field-induced effective magnetic field due to the
SOC. Both effects increase with the THz field strength but
decrease with the THz frequency.
The amplitude of the steady-state spin polarization increases
with effect (ii), but decreases with effect (i) according to the
Pauli paramagnetism. At small THz field strength (and/or large THz
frequency) the hot-electron effect is weak and effect (ii)
dominates. The amplitude of the steady-state spin polarization thus
increases (decreases) with the field strength (frequency). At large
THz field strength (low THz frequency), the hot-electron
effect becomes more important than effect (ii) and the amplitude
of the steady-state spin polarization decreases (increases) with the
THz field strength (frequency).

We also find that the THz field can strongly change the SRT
 due to the two effects addressed above.
Specifically, the hot-electron effect shortens the SRT via enhancement of
momentum scattering and inhomogeneous broadening for the system
in weak/intermediate scattering limit. Meanwhile, effect~(ii)
increases the SRT due to the blocking of the
 inhomogeneous broadening.
For small impurity densities at low temperature, when the THz field strength is
  small (and/or the THz frequency is large), the hot-electron effect
  is weak, and effect (ii) becomes dominant. The SRT thus
increases (decreases) with
  the THz field strength (frequency). At large THz field strength
  (small THz frequency) the hot-electron effect is more important than
  effect (ii). The SRT thus decreases (increases)
  with the THz field strength (frequency). However, for large impurity
densities or high temperatures, the enhancement of the hot-electron effect overcomes
the increase of the effect (ii). Consequently the SRT
 decreases (increases) with the THz field
  strength (frequency). We also discuss the SOC dependence of the SRT
  at large impurity densities where the hot-electron
  effect dominates. For small SOC, which is in the
  strong scattering regime, the SRT first increases
with the THz field strength due to enhancement of momentum
 scattering, then decreases with it due to enhancement of the
inhomogeneous broadening. For large SOC, which is
in the weak/intermediate scattering regime, increase of scattering also reduces
the SRT. Consequently, the SRT
decreases monotonically with the THz field strength.

\subsection{Discussion}
Finally we   compare our study with the
electric dipole spin resonance (EDSR)  in the literature.
To simplify the discussion, we introduce a simple spin
Hamiltonian which characterize the spin dynamics of our Hamiltonian
[Eq.~(\ref{hamilton})] and the EDSR:
\begin{equation}
  \hat{H}_{spin}(t) = \frac{1}{2}(\Delta_0 + \Delta_1 )
  \hat{\sigma}_x + \frac{1}{2} \Delta_2 \hat{\sigma}_y + \gamma
  \cos(\Omega t) \hat{\sigma}_y.
\end{equation}
Here $\Delta_1$ and $\Delta_2$ characterize the
  ${\bf k}$-dependent effective magnetic fields due to the SOC.
  $\Delta_0$ represents the external static magnetic field used in the
  EDSR set-up.\cite{Rashba0,Rashba1,Rashba2,Rashba3,Duckheim}
(For our case: $\Delta_0=0$, $\Delta_1=2\alpha_Rk_y$,
  $\Delta_2=-2\alpha_Rk_x$, $\gamma=-\alpha_R e E/\Omega$.)
 In EDSR, $\Delta_0$ is usually much larger than $\Delta_1$, $\Delta_2$ and
  $\gamma$, and $\Delta_0=\Omega$.\cite{Kato1,Meier} To the
  lowest order approximation, the spin dynamics is governed by
  $H^0_{spin}=\frac{1}{2}\Delta_0
\hat{\sigma}_x+\gamma\cos(\Omega t)\hat{\sigma}_y$.
 In the RWA the solutions of the Sch\"odinger
 equation are given by $\Psi_{\pm} = e^{\pm i \gamma t} \frac{1}{\sqrt{2}} ( e^{-i\Omega
  t/2} \chi_{+} \pm i e^{i\Omega t/2} \chi_{-}) $
with $\hat{\sigma}_x \chi_{\pm} = \pm \chi_{\pm}$. With initial
condition $\psi=\chi_{+}$, one obtains $\langle S_x \rangle=
\frac{1}{2} \cos(\gamma t)$, which is the well-known Rabi oscillation.
The ${\bf k}$-dependent $\Delta_1$ and $\Delta_2$ effective magnetic
fields lead to the damping of the Rabi oscillation due to the
DP mechanism in the
presence of scattering.\cite{Duckheim} For the case of strong driving field
($\gamma\gtrsim \Omega$), the solutions of the Sch\"odinger equation
are the Floquet wave functions $\Psi_{{\bf k}\eta}$ given in
Eq. (\ref{Floquet}). Now the spin dynamics is given by
\begin{eqnarray}
  \langle S_i \rangle &=&\sum_{{\bf
  k};\eta,\eta^{\prime};n,m;\sigma,\sigma^{\prime}} \rho^{F
(\eta^{\prime}\eta)}_{\bf k}
  \upsilon_{n\sigma}^{{\bf k}\eta\ast} \upsilon_{m\sigma^{\prime}}^{{\bf
  k}\eta^{\prime}} \langle \sigma| \frac{1}{2} \hat{\sigma}_{i} |\sigma^{\prime}
  \rangle \nonumber \\
 && \mbox{} \times e^{it[y_{{\bf
    k}\eta}-y_{{\bf k}\eta^{\prime}}+(m-n)\Omega ]},
\end{eqnarray}
with $i=x,y,z$. From the above equation, it is seen that, unlike the
weak driving-field case where only a single Rabi frequency is
observable, here the spin signal $S_i(t)$ oscillates at many frequencies
$y_{{\bf k}\eta}-y_{{\bf k}\eta^{\prime}}+(m-n)\Omega$ (with $m-n=0,\pm1,\pm2,\cdots$).
Moreover, in our case, $\Delta_1$, $\Delta_2$, $\gamma$ and $\Omega$ are on the
same order of magnitude while $\Delta_0=0$. Thus the spin precession
frequency varies largely with ${\bf k}$. This large inhomogeneous
broadening of spin precession frequency smears out the driving
field-induced Rabi oscillation of the spin polarization signals.

\begin{acknowledgments}
  This work was supported by the Natural Science Foundation of China
  under Grant Nos. 10574120 and 10725417,
the National Basic Research Program of China under Grant
No.\ 2006CB922005, and and the Innovation
  Project of Chinese Academy of Sciences.
  One of the authors (M.W.W.) was also partially supported by the
Robert-Bosch Stiftung and GRK 638. He would like to thank J. Fabian and
C. Sch\"uller at Universit\"at Regensburg and M. Aeschlimann at
Technische Universit\"at
Kaiserslautern for hospitality where part of this work was finalized. J.H.J.
would like to thank M. Q. Weng, J. L. Cheng and Y. Ji for helpful discussion.

\end{acknowledgments}

\begin{appendix}

\section{Derivation of electron-impurity scattering term
 in Floquet-Markov limit}

Here we  give terms due to electron-impurity scattering as an
example. Terms due to other scattering can be obtained similarly.
 From nonequilibrium Green function theory,\cite{Haugbook} the
electron-impurity scattering term can be written as
\begin{equation}
  \left. \partial_{t}\rho_{{\bf k}}\right|_{ei}
  = \Big\{- {\cal{A}}^{ei}_{{\bf k}}(><) + {\cal{A}}^{ei}_{{\bf k}}(<>) \Big\}
  + \Big\{...\Big\}^{\dagger},
\end{equation}
where
\begin{eqnarray}
 &&\hspace{-1.cm} {\cal{A}}^{ei}_{{\bf k}}(><) = \sum_{{\bf k}^{\prime},q_z}
  n_{i}U_{{\bf k}-{\bf k}^{\prime},q_z}^2|I(iq_z)|^2 \nonumber \\ &&
  \hspace{-0.cm} \mbox{} \times \int_{-\infty}^{t} \!\!d{\tau}
  \hat{U}_0^e({\bf k}^{\prime},t,\tau) \hat{\rho}^{>}_{{\bf
      k}^{\prime}}(\tau) \hat{\rho}^{<}_{\bf k}(\tau) \hat{U}_0^e({\bf k},\tau,t).
\end{eqnarray}
${\cal{A}}^{ei}_{{\bf k}}(<>)$ can be obtained by interchanging $>$ and $<$.
It is better to work in the interaction picture, or the ``Floquet
picture'':
\begin{equation}
  \hat{\rho}^{\gtrless F}_{\bf k}(t) = \hat{U}_0^{e\ \dagger}({\bf k},t,0)
  \hat{\rho}^{\gtrless}_{\bf k}(t) \hat{U}_0^{e}({\bf k},t,0).
\end{equation}
After this transformation, the term becomes
\begin{eqnarray}
   {\tilde{\cal{A}}^{ei}}_{{\bf k}}(><) &=&
  \hat{U}_0^{e\ \dagger}({\bf k},t,0) {\cal{A}}^{ei}_{{\bf k}}(><)
 \hat{U}_0^{e}({\bf k},t,0) \nonumber \\
 &=& \sum_{{\bf k}^{\prime},q_z}
  n_{i}U_{{\bf k}-{\bf k}^{\prime},q_z}^2|I(iq_z)|^2 \int_{-\infty}^{t} \!\!d{\tau}
  \hat{S}_{{\bf k},{\bf k}^{\prime}}(t,0) \nonumber \\
 &&\hspace{-0.cm}\mbox{}\times \hat{\rho}^{> F}_{{\bf
      k}^{\prime}}(\tau) \hat{S}_{{\bf k}^{\prime},{\bf
      k}}(\tau,0)  \hat{\rho}^{< F}_{\bf k}(\tau),
\end{eqnarray}
where $\hat{S}_{{\bf k},{\bf k}^{\prime}}(t,0) = \hat{U}_0^{e\
  \dagger}({\bf k},t,0) \hat{U}_0^{e}({\bf k}^{\prime},t,0)$.
According to Floquet-Markov theory, the Markov approximation should be
made with respect to the spectrum determined by the Floquet
wavefunctions, i.e., $\hat{\rho}^{\gtrless F}_{\bf k}(\tau)\approx
\hat{\rho}^{\gtrless F}_{\bf k}(t)$. Thus, the scattering term becomes
\begin{eqnarray}
 \hspace{-0.5cm} {\tilde{\cal{A}}^{ei}}_{{\bf k}}(><) &=&
  \sum_{{\bf k}^{\prime},q_z} n_{i}U_{{\bf k}-{\bf k}^{\prime},q_z}^2|I(iq_z)|^2
  \hat{S}_{{\bf k},{\bf k}^{\prime}}(t,0) \nonumber \\
&& \hspace{-0.cm}\mbox{}\times \hat{\rho}^{> F}_{{\bf k}^{\prime}}(t)
 \int_{-\infty}^{t} \!\!d{\tau} \hat{S}_{{\bf
      k}^{\prime},{\bf k}}(\tau,0) \hat{\rho}^{< F}_{\bf k}(t).
\end{eqnarray}
The next step toward the explicit form of the scattering term is based
on the analysis of the elements of $\hat{S}$. Expanding the kinetic
equations in the basis of \{$|\xi_{{\bf k}\eta}(0)\rangle$\},
the elements of $\hat{S}$ are given by
\begin{eqnarray}
 S^{(\eta_1\eta_2)}_{{\bf k}^{\prime},{\bf
      k}}(t,0)&=&\langle \xi_{{\bf
    k}^{\prime}\eta_1}(t)|\xi_{{\bf k}\eta_2}(t)\rangle \nonumber \\
&& \mbox{}\times
e^{i[(\varepsilon_{{\bf k}^{\prime}}-\varepsilon_{\bf
      k})t+\gamma_E\sin(\Omega t)(k_x^{\prime}-k_x)]} \nonumber \\
 &=& \sum_{n} S^{(n)(\eta_1\eta_2)}_{{\bf
    k}^{\prime},{\bf k}}
e^{i t(n\Omega+\bar{\varepsilon}_{{\bf
      k}^{\prime}\eta_1}-\bar{\varepsilon}_{{\bf k}\eta_2}) }.
\end{eqnarray}
The scattering term can then be explicitly laid out by expanding all
the operators in the basis of \{$|\xi_{{\bf k}\eta}(0)\rangle$\}:
\begin{eqnarray}
 \hspace{0.15cm} {\tilde{\cal{A}}^{ei}}_{{\bf
  k}}(><)|^{(\eta\eta^{\prime})} &=& \sum_{{\bf
  k}^{\prime},q_z,n,\eta_1\eta_2\eta_3} \pi n_i U_{{\bf k}-{\bf
  k}^{\prime},q_z}^2 |I(iq_z)|^2 \nonumber \\
  && \hspace{-1.6cm}\mbox{} \times S^{(\eta\eta_1)}_{{\bf k},{\bf
  k}^{\prime}}(t,0)  \rho^{> F(\eta_1\eta_2)}_{{\bf
      k}^{\prime}}(t) S^{(n)(\eta_2\eta_3)}_{{\bf k}^{\prime},{\bf k}}  \rho^{<
  F(\eta_3\eta^{\prime})}_{\bf k}(t)
  \nonumber \\
&& \hspace{-1.6cm}\mbox{} \times
\delta(n\Omega+\bar{\varepsilon}_{{\bf k}^{\prime}\eta_2}-\bar{\varepsilon}_{{\bf
  k}\eta_3}).
\end{eqnarray}
According to Eq.\ (A1), one can readily arrive at the full expression
of the electron-impurity scattering term, which is exactly Eq.\ (\ref{scatei}).

\section{Numerical scheme}

Our numerical scheme is based on the scheme laid out in detail in Ref.\
\onlinecite{wu-hot-e}, where the nonlinear kinetic spin Bloch equations
are solved self-consistently with high
accuracy.\cite{wu-later,wu-lowT,wu-Exp-HighP,wu-Exp-Aniso}
The scheme is based on a discretization of the two dimensional momentum
space with $N\times M$ control regions where the ${\bf k}$-grid points
are chosen to be ${\bf k}_{l,m}=\sqrt{2m^{\ast}E_{l}}(\cos \theta_m,
\sin \theta_m)$. In principle, the coherent terms are easily
solved. However, the scattering terms are difficult to solve as the
$\delta$-functions are hard to be integrated numerically.
To facilitate the evaluation of the $\delta$-functions in the
scattering terms, we set $E_{l}=(l+1/2)\Delta E$ and
$\omega_{LO}=n_{LO}\Delta E$ where $l$ and $n_{LO}$ are integer
numbers and $\Delta E$ is the energy span in each control
region.\cite{wu-hot-e} To apply this scheme to the  kinetic
spin Bloch equations with THz field, we also set $\Omega=n_{\mbox{\tiny THz}}
\Delta E$ with
$n_{\mbox{\tiny THz}}$ being integer number (typically $1\sim 3$ in our
calculation). However, the $\delta$-functions are still difficult to be
evaluated as $y_{{\bf k}\eta}$ and $\Delta E$ are not
commensurable.
We therefore use the approximation  $y_{{\bf k}\eta} \approx N^y_{{\bf
    k}\eta} \Delta E $, with $N^y_{{\bf k}\eta}$ being the integer
part of $y_{{\bf k}\eta}/\Delta E$. This approximation affects the spin
kinetics marginally as $|y_{{\bf k}\eta}-N^y_{{\bf k}\eta}\Delta E|$ is
usually much smaller than  $k_BT$ and/or the chemical
potential. Moreover, as the driving field is very strong, the spectrum
of the Floquet states is mainly determined by the sideband
effect and $y_{{\bf k}\eta}-N^y_{{\bf k}\eta}\Delta E$  only
plays a quite marginal role. Furthermore, one can approach the exact
results by increasing $n_{\mbox{\tiny THz}}$. In our computation,
we make sure that for $n_{\mbox{\tiny THz}}$
we choose, the relative error is less than $5$\ \%.
To make the treatment consistent, we also approximate
$S^{(\eta_1\eta_2)}_{{\bf k}^{\prime},{\bf k}} (t,0) \approx \sum_{n}
S^{(n)(\eta_1\eta_2)}_{{\bf k}^{\prime},{\bf k}}
e^{it[n\Omega+\varepsilon_{{\bf k}^{\prime}}-\varepsilon_{\bf k} +
    (N^y_{{\bf k}^{\prime}\eta_1}-N^y_{{\bf k}\eta_2})\Delta E]}$.
Or more concisely, $\hat{S}_{{\bf k}^{\prime},{\bf k}} (t,0) \approx \sum_{n}
\hat{{\cal{R}}}^{(n)}_{{\bf k}^{\prime},{\bf k}}
e^{it(n\Delta E+\varepsilon_{{\bf k}^{\prime}}-\varepsilon_{\bf
      k})}=\hat{{\cal{R}}}_{{\bf k}^{\prime},{\bf k}}(t,0)$,
where ${\cal{R}}^{(n)(\eta_1\eta_2)}_{{\bf k}^{\prime},{\bf
    k}}=S^{(m)(\eta_1\eta_2)}_{{\bf k}^{\prime},{\bf k}}$ with $m$
satisfying $mn_{\mbox{\tiny THz}}+ N^y_{{\bf k}^{\prime}\eta_1}-N^y_{{\bf
    k}\eta_2}=n$. Correspondingly, $\hat{T}_{{\bf k}^{\prime},{\bf k}}
(t,0) \approx \sum_{n} \hat{{\cal{W}}}^{(n)}_{{\bf k}^{\prime},{\bf k}}
e^{it(n\Delta E+\varepsilon_{{\bf k}^{\prime}}-\varepsilon_{\bf
      k})}=\hat{{\cal{W}}}_{{\bf k}^{\prime},{\bf k}}(t,0)$,
with ${\cal{W}}^{(n)(\eta_1\eta_2)}_{{\bf k}^{\prime},{\bf
    k}}=T^{(m)(\eta_1\eta_2)}_{{\bf k}^{\prime},{\bf k}}$.
We keep the coherent precession due to $y_{{\bf k}\eta}-N^y_{{\bf
    k}\eta}\Delta E$  by adding it into the coherent term.
After these approximations, the coherent and scattering terms of
the kinetic spin Bloch equations read
\begin{eqnarray}
&&\hspace{-0.65cm} \left. \partial_{t}\hat{\rho}^{F}_{{\bf k}}(t)\right|_{coh} =
   i\Big[\sum_{{\bf k}^{\prime},q_z,n} V_{{\bf k}-{\bf
      k}^{\prime},q_z}|I(iq_z)|^2 \hat{{\cal{W}}}_{{\bf k},{\bf
      k}^{\prime}}(t,0)\hat{\rho}^{F}_{{\bf k}^{\prime}}(t) \nonumber \\
&&\hspace{1.4cm} \mbox{} \times \hat{{\cal{W}}}_{{\bf k}^{\prime},{\bf
    k}}(t,0)-\hat{H}_{r}({\bf k}), \  \hat{\rho}^{F}_{{\bf k}}(t)\Big],
\end{eqnarray}
with $H_{r}({\bf k})^{\eta_1,\eta_2}=\delta_{\eta_1,\eta_2}(y_{{\bf k}\eta_1}-N^y_{{\bf
    k}\eta_1}\Delta E)$,
\begin{eqnarray}
\hspace{-0.2cm}\left.\partial_{t}\hat{\rho}^{F}_{{\bf k}}(t)\right|_{ei} &=&
- \sum_{{\bf k}^{\prime},n,q_z} \pi n_{i} U^2_{{\bf k}-{\bf
      k}^{\prime},q_z} |I(iq_z)|^2 \nonumber \\ &&
      \hspace{-1.cm} \mbox{} \times \delta(n\Delta E+\varepsilon_{{\bf
  k}^{\prime}}-\varepsilon_{\bf k}) \bigg{[} \Big{\{}
\hat{{\cal{R}}}_{{\bf k},{\bf k}^{\prime}}(t,0)\hat{{\cal{R}}}^{(n)}_{{\bf
      k}^{\prime},{\bf k}} \hat{\rho}^{F}_{\bf k}(t) \nonumber \\ &&
      \hspace{-1.cm}\mbox{}
 - \hat{{\cal{R}}}_{{\bf k},{\bf k}^{\prime}}(t,0)
  \hat{\rho}^{F}_{{\bf k}^{\prime}}(t)\hat{{\cal{R}}}^{(n)}_{{\bf
      k}^{\prime},{\bf k}}\Big{\}}\ + \Big{\{}...\Big{\}}^{\dagger} \bigg{]},
\end{eqnarray}
\begin{eqnarray}
\left.\partial_{t}\hat{\rho}^{F}_{{\bf k}}(t)\right|_{ep} &=& -
      \sum_{{\bf k}^{\prime},n,\lambda,\pm,q_z} \pi |M_{\lambda, {\bf
      k}-{\bf k}^{\prime},q_z}|^2 |I(iq_z)|^2 \nonumber \\ &&
      \hspace{-1.2cm} \mbox{} \times
      \delta(\pm\omega_{\lambda, {\bf k}-{\bf k}^{\prime}, q_z}
  +n\Delta E+\varepsilon_{{\bf k}^{\prime}}-\varepsilon_{\bf
  k}) e^{\mp it\omega_{\lambda, {\bf k}-{\bf k}^{\prime}, q_z}}
\nonumber \\ && \hspace{-1.2cm} \mbox{} \times
  \bigg{[} \Big{\{}  N^{\pm}_{\lambda, {\bf k}-{\bf k}^{\prime}, q_z}
  \hat{{\cal{R}}}_{{\bf k},{\bf k}^{\prime}}(t,0)
  \big{(}\hat{1}-\hat{\rho}^{F}_{{\bf k}^{\prime}}(t) \big{)}
  \hat{{\cal{R}}}^{(n)}_{{\bf k}^{\prime},{\bf k}} \hat{\rho}^{F}_{\bf
      k}(t) \nonumber \\ && \hspace{-1.7cm} \mbox{}
  - \hat{{\cal{R}}}_{{\bf k},{\bf k}^{\prime}}(t,0)
  \hat{\rho}^{F}_{{\bf k}^{\prime}}(t)\hat{{\cal{R}}}^{(n)}_{{\bf
      k}^{\prime},{\bf k}} \big{(}\hat{1}-\hat{\rho}^{F}_{{\bf
      k}}(t) \big{)}
  \Big{\}}
 + \Big{\{}...\Big{\}}^{\dagger} \bigg{]} , %\nonumber \\ && \hspace{1.2cm}
\end{eqnarray}
\begin{eqnarray}
\left.\partial_{t}\hat{\rho}^{F}_{{\bf k}}(t)\right|_{ee} &=& -
\sum_{{\bf k}^{\prime},{\bf k}^{\prime\prime},n,n^{\prime}} \pi
 \Big{[}\sum_{q_z} V_{{\bf k}-{\bf k}^{\prime}, q_z } |I(iq_z)|^2
 \Big{]}^2 \nonumber \\ &&
  \hspace{-1.45cm} \mbox{} \times \delta(n \Delta E+\varepsilon_{{\bf
  k}^{\prime}}-\varepsilon_{\bf k}+\varepsilon_{{\bf
  k}^{\prime\prime}}  -\varepsilon_{{\bf k}^{\prime\prime}-{\bf
  k}+{\bf k}^{\prime}})
\nonumber \\ && \hspace{-1.45cm} \mbox{} \times
  \bigg{[} \Big{\{} \hat{{\cal{W}}}_{{\bf k},{\bf k}^{\prime}}(t,0)
    \big{(}\hat{1}-\hat{\rho}^{F}_{{\bf k}^{\prime}}
  (t)\big{)}\hat{{\cal{W}}}^{(n^{\prime})}_{{\bf k}^{\prime},{\bf
  k}}\hat{\rho}^{F}_{\bf k}(t) \nonumber \\ &&
  \hspace{-1.2cm}\mbox{} \times
    \mbox{Tr}\big{[}\hat{{\cal{W}}}^{(n-n^{\prime})}_{{\bf k}^{\prime\prime},{\bf
        k}^{\prime\prime}-{\bf k}+{\bf k}^{\prime}} \hat{\rho}^{F}_{{\bf
        k}^{\prime\prime}-{\bf k}+{\bf k}^{\prime}}(t) \hat{{\cal{W}}}_{{\bf
        k}^{\prime\prime}-{\bf k}+{\bf k}^{\prime},{\bf
        k}^{\prime\prime}}(t,0)
\nonumber \\ &&
  \hspace{-1.2cm} \mbox{} \times
    \big{(}\hat{1}-\hat{\rho}^{F}_{{\bf k}^{\prime\prime}}(t)\big{)}\big{]}
    \ - \
   \hat{{\cal{W}}}_{{\bf k},{\bf k}^{\prime}}(t,0) \hat{\rho}^{F}_{{\bf k}^{\prime}}(t)
\hat{{\cal{W}}}^{(n^{\prime})}_{{\bf k}^{\prime},{\bf
    k}} \nonumber \\ &&
  \hspace{-1.2cm}\mbox{} \times
\big{(}\hat{1}-\hat{\rho}^{F}_{\bf k}(t)\big{)}
    \mbox{Tr}\big{[}\hat{{\cal{W}}}^{(n-n^{\prime})}_{{\bf
      k}^{\prime\prime},{\bf k}^{\prime\prime}
-{\bf k}+{\bf k}^{\prime}} \big{(}\hat{1}-\hat{\rho}^{F}_{{\bf
   k}^{\prime\prime}-{\bf k}+{\bf k}^{\prime}}(t)\big{)}
\nonumber \\ &&
  \hspace{-1.2cm}\mbox{} \times
   \hat{{\cal{W}}}_{{\bf k}^{\prime\prime}-{\bf k}+{\bf k}^{\prime},{\bf
        k}^{\prime\prime}}(t,0)
\hat{\rho}^{F}_{{\bf k}^{\prime\prime}}(t)
\big{]}\Big{\}}
    +\Big{\{}...\Big{\}}^{\dagger} \bigg{]}.
\end{eqnarray}

Now, the kinetic spin Bloch equations can be treated via the numerical
scheme in Ref.\ \onlinecite{wu-hot-e}. The only difference is the
summations over sideband indexes which increase the complexity of the
calculation. Typically, the sideband index runs through
$[-24,24]$ ($[-3,3]$) for the electron-impurity and electron-phonon
scattering (the electron-electron Coulomb scattering) to converge the
results when the THz field is $E=1.5$~kV/cm with $\nu=0.65$~THz.

\begin{figure}[htb]
\includegraphics[height=4.3cm]{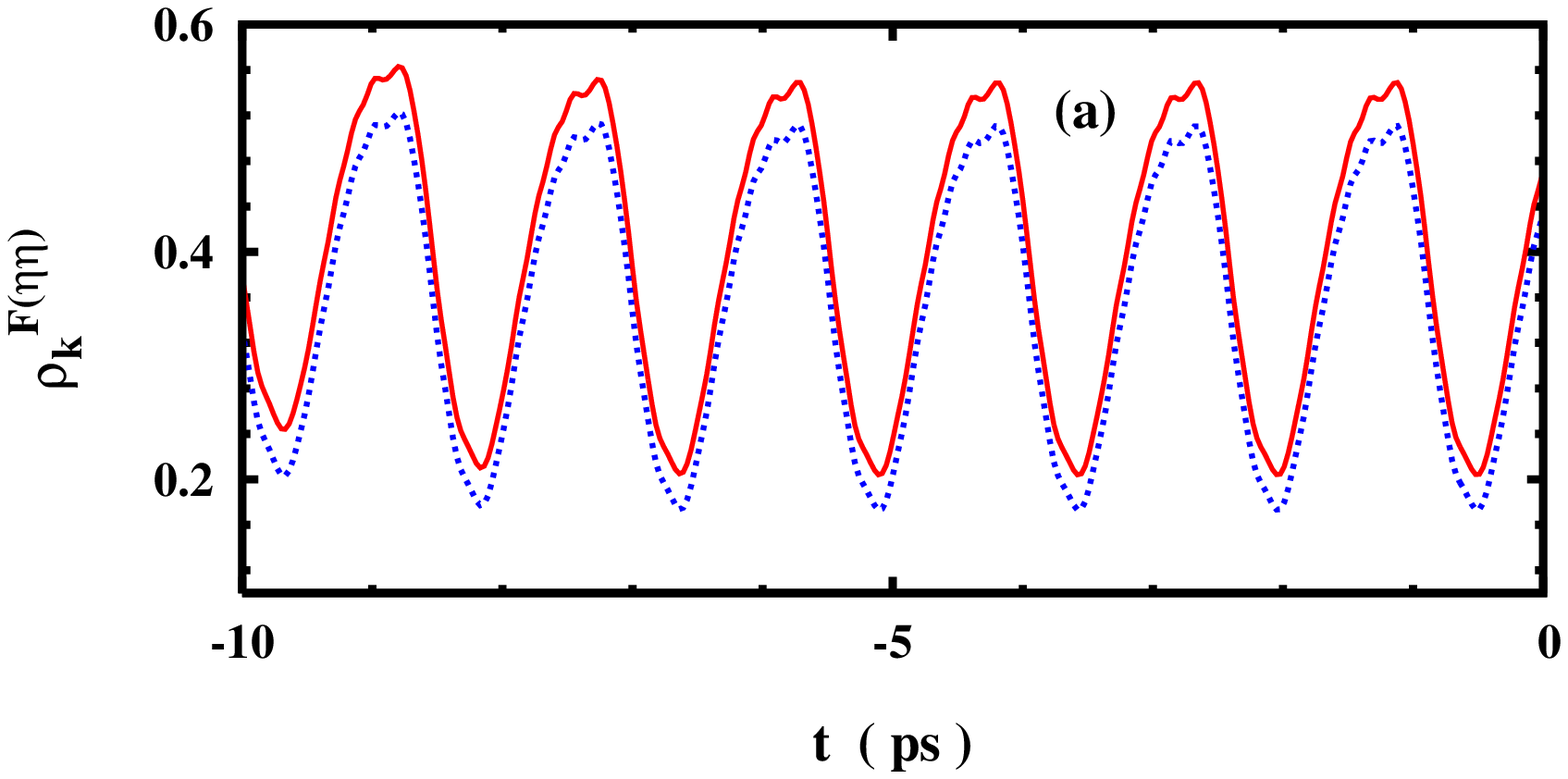}
\includegraphics[height=4.3cm]{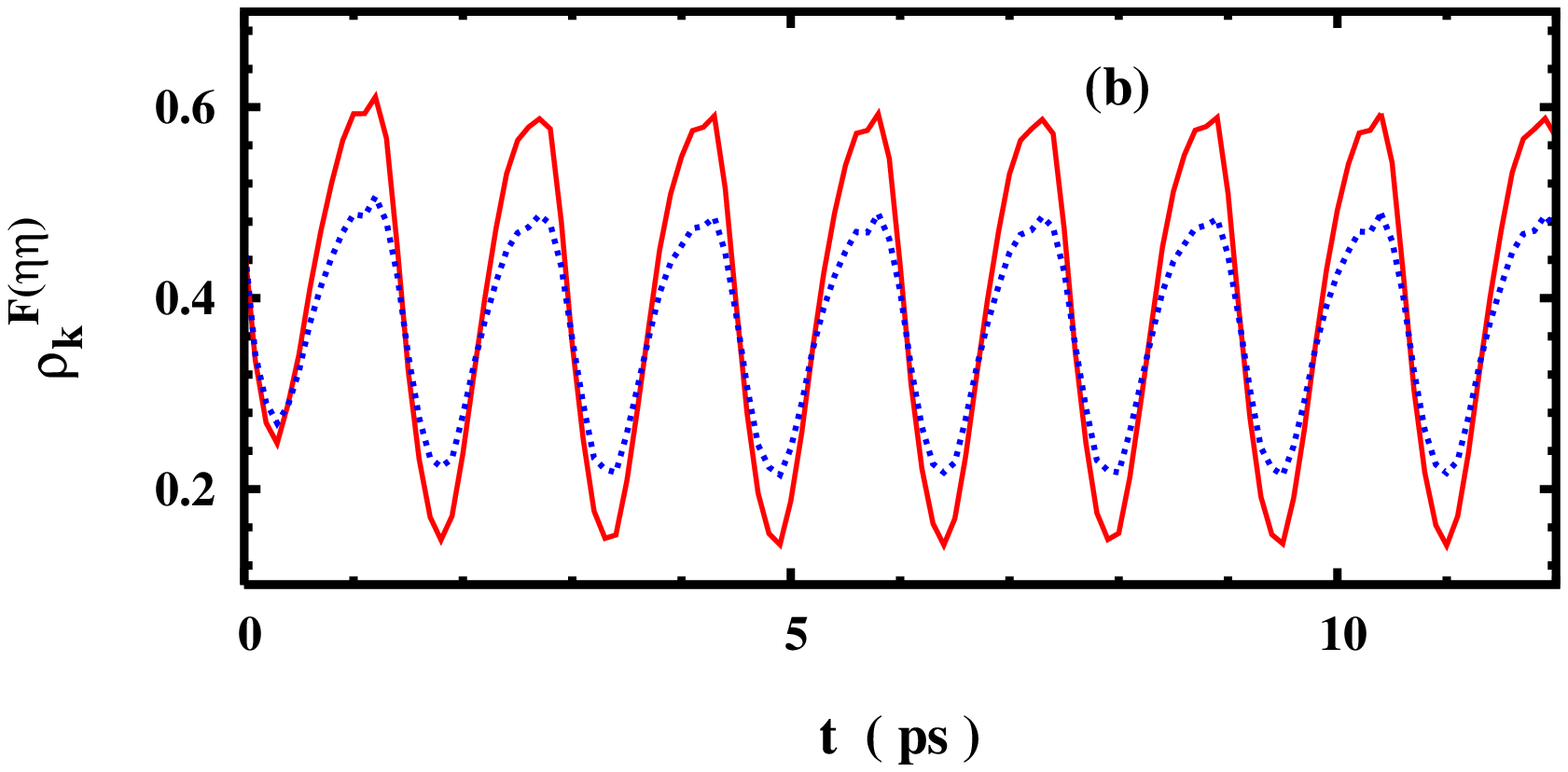}
\caption{(Color online) Time evolution of the distribution of two
Floquet states with ${\bf k}=(1.1k_F^{0},0)$: $\rho_{\bf k}^{F(--)}$
  (solid curve) and $\rho_{\bf k}^{F(++)}$ (dotted curve). (a):
initial distribution preparation without the SOC; (b): after the
preparation with the SOC. $E=1.5$\ kV/cm, $T=50$\ K and $N_i=0.05 N_e$.}
\label{fig:dis_evo}
\end{figure}

As mentioned in Sec.\ III, the initial distribution of the electron system
is chosen to be a spin polarized hot-electron distribution under the
THz field, which is obtained by sufficient long time (typically
$\sim10$~ps) evolution from a spin-polarized Fermi distribution at the
lattice temperature with the SOC being turned off.\cite{wu-hot-e} In Fig.\
\ref{fig:dis_evo}, we plot the time evolution of the distribution on
the two Floquet states with ${\bf k}=(1.1k_F^{0},0)$ ($k_F^{0}$ is the
Fermi wave vector): $\rho_{\bf k}^{F(--)}$ (solid curve) and $\rho_{\bf
  k}^{F(++)}$ (dotted curve) for the case with 4\ \% initial spin
polarization along the $z$ axis.

It is seen from Fig.\ \ref{fig:dis_evo}(a) that after only
about $3$\ ps, the distributions show regular
oscillations. This indicates that the system reaches its steady
state. Moreover, the periods of the oscillations are close to the
period of the THz field $T_0$. As we have pointed out in Sec.\ II, the
eigen-modes of the steady-state distributions have the
general form $\hat{\tilde{\rho}}^{\alpha}_{\bf k}=e^{i\mu_{\bf
  k}^{\alpha}t}\sum_{n}\hat{Q}_{{\bf k}}^{\alpha,n}e^{in\Omega t}$
($\alpha=1,2,3,4$) according to the Floquet theorem.\cite{ACReview2}
The two diagonal elements of the distribution function should be
governed by one of these modes while other modes are damping modes
which do not appear in the steady state. When the THz field is not too
large, the eigen-values of the relevant eigen-modes $\mu_{\bf
  k}^{\alpha}$ are close to zero. %as it is zero in the limit of zero
%THz field.
Thus the distribution functions still have good periodic
behavior and the period is close to $T_0$.

\begin{figure}[htb]
\includegraphics[height=5.5cm]{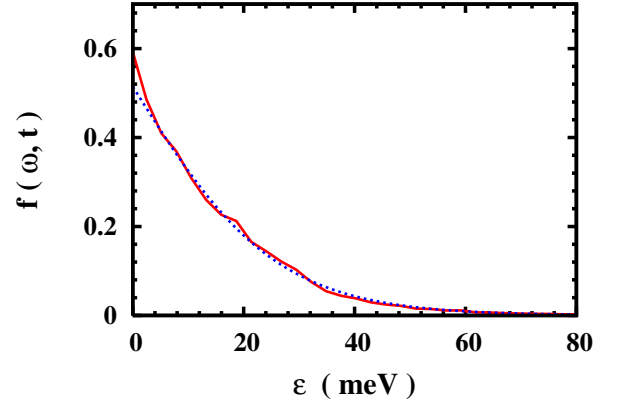}
\caption{(Color online) The hot-electron distribution in energy space
at $t=23$~ps (solid curve) and the fitting curve (dashed curve).
  $E=1.5$\ kV/cm, $T=50$\ K, and $N_i=0.05N_e$.}
\label{fig:disDOS}
\end{figure}

\begin{figure}[htb]
\includegraphics[height=6.cm]{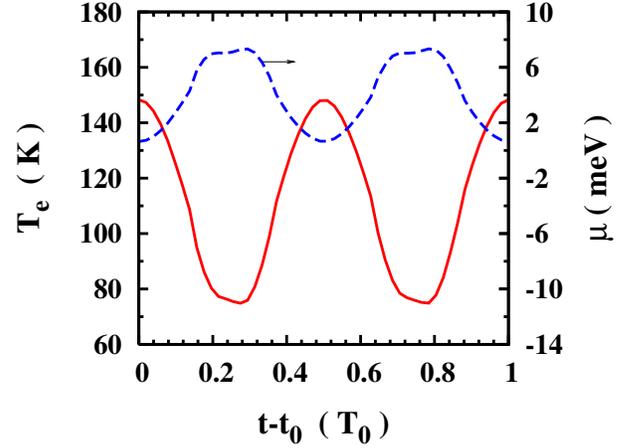}
\caption{(Color online) The hot-electron temperature $T_e$
  (solid curves) and the chemical potential $\mu$ (dashed curves) as
  function of $t$.
  $E=1.5$\ kV/cm, $T=50$\ K and $N_i=0.05N_e$.
  Note that the scale for the dashed curve is on the right hand
  side of the frame.}
\label{fig:Temu}
\end{figure}

From Fig.\
\ref{fig:dis_evo}(b) one further finds that when the SOC is included,
the steady-state distributions
still have good periodic behavior. The system
approaches steady state within 3\ ps and the period is again close to
$T_0$. The distribution difference on the two Floquet states in
Fig.~\ref{fig:dis_evo}(a) is due to the spin polarization  whereas in
Fig.\ \ref{fig:dis_evo}(b)  is caused by the spectral difference of
the two Floquet states.

\section{Hot-electron effect and hot-electron temperature}

As has been shown in Appendix\ B that the steady-state distribution
$\hat{\tilde{\rho}}^{F}_{{\bf k}}(t)$ is a time dependent function which
still exhibits good periodicity in our parameter regime. We can
extract the distribution in energy space at any time $t$ via Fourier
transformation:
\begin{equation}
  \hat{F}(\omega,t) \hat{D}(\omega,t) = \sum_{{\bf k}\eta}
    \tilde{\rho}^{F(\eta\eta)}_{{\bf k}} (t) \hat{D}_{{\bf
    k}\eta}(\omega,t).
\label{hotdis}
\end{equation}
Here $\hat{D}(\omega,t) = \sum_{{\bf k}\eta} \hat{D}_{{\bf
    k}\eta}(\omega,t)$ is the generalized density of states
($2\times2$ matrix) where $t$ is the center-of-mass time:\cite{Jauho,Cheng,Jiang}
\begin{equation}
\hat{D}_{{\bf k}\eta}(\omega,t) = \int_{-\infty}^{\infty}
\!\!\frac{d{\tau}}{2\pi} e^{i\omega \tau} \Psi_{{\bf k}\eta}
(t+\frac{\tau}{2}) \Psi_{{\bf k}\eta}^{\dagger} (t-\frac{\tau}{2}).
\end{equation}
It has been found that $\hat{D}_{{\bf k}\eta}(\omega,t)$ and
$\hat{D}(\omega,t)$ are periodic functions of $t$ with the same period
as that of the THz field $T_0=2\pi/\Omega$. Therefore, the
distribution $\hat{F}(\omega,t)$ is also a periodic function of
$t$ with period $T_0$. According to the symmetry analysis,
$\hat{D}(\omega,t)=D_{\uparrow\uparrow}(\omega,t)\hat{1}-\mbox{Im}
\{D_{\uparrow\downarrow}(\omega,t)\}\hat{\sigma}_{y}$.\cite{Cheng}
As the matrices $\hat{1}$ and $\hat{\sigma}_{y}$ form a group, the
distribution $\hat{F}(\omega,t)$ should also be decomposed into two
parts:
$\hat{F}(\omega,t)=f(\omega,t)\hat{1}+s(\omega,t)\hat{\sigma}_{y}$.
Equation\ (\ref{hotdis}) then turns into [denoting $\hat{\zeta}(\omega,t)= \sum_{{\bf k}\eta}
    \tilde{\rho}^{F(\eta\eta)}_{{\bf k}}(t) \hat{D}_{{\bf
    k}\eta}(\omega,t)$]:
\begin{eqnarray}
&&  f(\omega,{t}) D_{\uparrow\uparrow}(\omega,{t}) - s(\omega,{t})
    \mbox{Im}\{D_{\uparrow\downarrow}(\omega,{t})\} \nonumber \\
&&= \zeta^{\uparrow\uparrow}(\omega,{t}), \\
&&\mbox{}  - s(\omega,{t}) D_{\uparrow\uparrow}(\omega,{t}) + f(\omega,{t})
    \mbox{Im}\{D_{\uparrow\downarrow}(\omega,{t})\} \nonumber \\
&& = \mbox{Im}\{\zeta^{\uparrow\downarrow}(\omega,{t})\}.
\end{eqnarray}
The solutions of the above equations are given by
\begin{eqnarray}
\hspace{-0.5cm} f &=& \frac{ \zeta^{\uparrow\uparrow} D_{\uparrow\uparrow} -
  \mbox{Im}\{\zeta^{\uparrow\downarrow}\}
    \mbox{Im}\{D_{\uparrow\downarrow}\} } {
    D_{\uparrow\uparrow}^2 -
    \big{[}\mbox{Im}\{D_{\uparrow\downarrow}\}\big{]}^2  }, \\
\hspace{-0.5cm} s &=& \frac{ - \mbox{Im}\{ \zeta^{\uparrow\downarrow} \}
    D_{\uparrow\uparrow} +  \zeta^{\uparrow\uparrow}
    \mbox{Im}\{D_{\uparrow\downarrow}\} } {D_{\uparrow\uparrow}^2 -
    \big{[}\mbox{Im}\{D_{\uparrow\downarrow}\}\big{]}^2  }.
\end{eqnarray}

One notices that Eq.\ (C1) is a natural generalization of the distribution in
energy space from thermal equilibrium to the nonequilibrium case.
It is straightforward to see that in the zero THz field limit the distribution
$\hat{F}(\omega,t)$ recovers the Fermi distribution as
 $\hat{D}_{{\bf   k}\eta}(\omega,{t})=\delta(\omega-\bar{\varepsilon}_{{\bf
    k}\eta})|\xi_{{\bf k}\eta}\rangle\langle\xi_{{\bf k}\eta}|$ and $\tilde{\rho}^{F(\eta\eta)}_{{\bf k}} (t) =
f_{F}(\bar{\varepsilon}_{{\bf k}\eta})$ in the zero-field
limit ($f_{F}(x)$ is the Fermi distribution function). Therefore
from Eq.\ (C1), $\hat{F}(\omega,t)= f_{F}(\omega)\hat{1}$.

A typical $f(\omega,t)$ is plotted in Fig.\ \ref{fig:disDOS}.
We use the hot-electron temperature $T_e$ to measure the hot-electron
effect. The hot-electron temperature is determined by fitting the tail
of $f(\omega,t)$ with Fermi distribution function. We plot the
fitted hot-electron temperature $T_e$ and the chemical potential $\mu$
in Fig.\ \ref{fig:Temu} ($t_0$ in the
figure denotes the starting time which is 21.5\ ps). It is seen in Fig.~\ref{fig:Temu} that
$T_e$ and $\mu$ are also periodic functions of $t$ with periodicity
of $T_0/2$. This is because these quantities only depend on the
strength of the THz field. The hot-electron temperature used in
Sec.\ III is the largest temperature, which is sufficient in measuring the
hot-electron effect.

\end{appendix}

\end{document}